\newcommand*{\ATLASLATEXPATH}{latex/}

\documentclass[cernpreprint,texlive=2011,txfonts,UKenglish]{\ATLASLATEXPATH atlasdoc}
 \pdfoutput=1
\usepackage[siunitx=false]{\ATLASLATEXPATH atlaspackage}
\usepackage{\ATLASLATEXPATH atlasphysics}
\graphicspath{{logos/}{figures/}}

\usepackage{graphicx}
\usepackage{booktabs}
\usepackage{cite}

\usepackage{amsmath}
\usepackage{amssymb}
\usepackage{xspace}
\usepackage{epstopdf}
\usepackage{microtype}
\usepackage{multirow}
\usepackage{verbatim}
\usepackage[absolute,overlay]{textpos}

\newcommand{\mlead}{\ensuremath{m_{\mathrm{2j}}^{\mathrm{lead}}}\xspace}
\newcommand{\msublead}{\ensuremath{m_{\mathrm{2j}}^{\mathrm{subl}}}\xspace}
\newcommand{\mfourj}{\ensuremath{m_{\mathrm{4j}}}\xspace}
\newcommand{\mtwoJ}{\ensuremath{m_{\mathrm{2J}}}\xspace}

\def\Grav{\ensuremath{G^{*}_{\mathrm{KK}}}\xspace}
\def\mGrav{\ensuremath{m_{G^{*}_{\mathrm{KK}}}}\xspace}
\def\hbb{\ensuremath{h\hspace{-0.4 mm}\to\hspace{-0.4 mm}b\bar{b}}\xspace}
\newcommand{\cosba}{\ensuremath{\cos\left(\beta-\alpha\right)}\xspace}
\newcommand{\tanb}{\ensuremath{\tan\beta}\xspace}
\newcommand{\dEta}{\ensuremath{\left|\Delta\eta_{\mathrm{dijets}}\right|}\xspace}
\def\accEff{\ensuremath{A\times\varepsilon}\xspace}
\def\Gtohhb{\Grav\ensuremath{\hspace{-0.4mm}\to\hspace{-0.4mm}hh\hspace{-0.4mm}\to\hspace{-0.4mm}b\bar{b}b\bar{b}}\xspace}
\def\Htohhb{\ensuremath{H\hspace{-0.4mm}\to\hspace{-0.4mm}hh\hspace{-0.4mm}\to\hspace{-0.4mm}b\bar{b}b\bar{b}}\xspace}
\def\pptoHtofourb{\ensuremath{pp\to\Htohhb}\xspace}
\def\pptoGtofourb{\ensuremath{pp\to\Gtohhb}\xspace}
\def\pptohh{$pp\hspace{-0.4mm}\to\hspace{-0.4mm} hh$\xspace}
\def\pptofourb{$pp\hspace{-0.4mm}\to\hspace{-0.4mm} hh\hspace{-0.4mm}\to\hspace{-0.4mm} b\bar{b}b\bar{b}$\xspace}

\def\sigGfourb{$\sigma\left(pp\hspace{-0.4mm}\to\hspace{-0.4mm}\Grav\hspace{-0.4mm}\to\hspace{-0.4mm}hh\hspace{-0.4mm}\to\hspace{-0.4mm}b\bar{b}b\bar{b}\right)$\xspace}
\newcommand\sigXfourb{\ensuremath{\sigma\left(pp\hspace{-0.4mm}\to\hspace{-0.4mm}X\hspace{-0.4mm}\to\hspace{-0.4mm}hh\hspace{-0.4mm}\to\hspace{-0.4mm}b\bar{b}b\bar{b}\right)}\xspace}

\def\ptlead{\ensuremath{\pt^{\mathrm{lead}}}\xspace}
\def\ptsubl{\ensuremath{\pt^{\mathrm{subl}}}\xspace}

\def\ttbar{$t\bar{t}$\xspace}
\def\bbbar{$b\bar{b}$\xspace}
\def\fourb{$b\bar{b}b\bar{b}$\xspace}

\def\akt{${\mathrm{anti-}}k_{t}$\xspace}
\def\pythia{{\sc Pythia}\xspace}
\def\powheg{{\sc Powheg}\xspace}
\def\powhegbox{{\sc Powheg-box}\xspace}
\def\madgraph{{\sc Madgraph}\xspace}
\def\sherpa{{\sc Sherpa}\xspace}
\def\dr{$\Delta R$\xspace}
\def\cls{$CL_{s}$\xspace}
\def\bbgg{\ensuremath{b\bar{b}\gamma\gamma}\xspace}

\def\alphatt{$\alpha_{t\bar{t}}$\xspace}
\newcommand\muqcd{\ensuremath{\mu_{\mathrm{QCD}}}\xspace}
\newcommand\kMPl{\ensuremath{k/\bar{M}_{\mathrm{Pl}}}\xspace}
\newcommand\frackMPl{\ensuremath{\frac{k}{\bar{M}_{\mathrm{Pl}}}}\xspace}

\hypersetup{pdftitle={ATLAS draft},pdfauthor={The ATLAS Collaboration}}
\AtlasTitle{Search for Higgs boson pair production in the \fourb final state from $pp$ collisions at $\sqrt{s} = 8$ \TeV{} with the ATLAS detector}
 \AtlasAbstract{
A search for Higgs boson pair production \pptohh is performed with 19.5\,fb$^{-1}$ of proton--proton collision data at $\sqrt{s}=8$\,\TeV, which were recorded by the ATLAS detector at the Large Hadron Collider in 2012. The decay products of each Higgs boson are reconstructed as a high-momentum \bbbar system with either a pair of small-radius jets or a single large-radius jet, the latter exploiting jet substructure techniques and associated $b$-tagged track-jets. No evidence for resonant or non-resonant Higgs boson pair production is observed. The data are interpreted in the context of the Randall--Sundrum model with a warped extra dimension as well as the two-Higgs-doublet model. An upper limit on the cross-section for \pptoGtofourb of 3.2\,(2.3) fb is set for a Kaluza--Klein graviton \Grav mass of 1.0 (1.5)\,\TeV, at the 95\% confidence level. The search for non-resonant Standard Model $hh$ production sets an observed 95\% confidence level upper limit on the production cross-section $\sigma(pp\hspace{-0.4mm}\rightarrow\hspace{-0.4mm}hh\hspace{-0.4mm}\rightarrow\hspace{-0.4mm}b\bar{b}b\bar{b})$ of 202\,fb, compared to a Standard Model prediction of $\sigma(pp\hspace{-0.4mm}\rightarrow\hspace{-0.4mm}hh\hspace{-0.4mm}\rightarrow\hspace{-0.4mm}b\bar{b}b\bar{b}) = 3.6 \pm 0.5$\,fb.
}
\AtlasJournal{Eur. Phys. J. C.}
\PreprintIdNumber{CERN-PH-EP-2015-099}

\begin{document}

 \maketitle

%\linenumbers

\section{Introduction}
%sec-introduction.tex
The discovery of a Higgs boson ($h$) \cite{Aad:2012tfa,Chatrchyan201230} at the Large Hadron Collider (LHC) consistent with the predictions of the Standard Model (SM) \cite{ATLAS:HiggsCouplings,PhysRevD.89.092007} motivates an enhanced effort to search for new physics via the Higgs sector. Many new physics models predict rates of Higgs boson pair production significantly higher than the SM rate~\cite{PhysRevD.58.115012,Grigo20131,PhysRevLett.111.201801}. 
For example, \TeV-scale resonances such as the first Kaluza-Klein (KK) excitation of the graviton,
\Grav, predicted in the bulk Randall-Sundrum (RS) model~\cite{Agashe:2007zd,Fitzpatrick} or the heavy neutral scalar, $H$, of two-Higgs-doublet models (2HDM)~\cite{Branco:2011iw} can decay into pairs of  Higgs bosons, $hh$. Enhanced non-resonant \pptohh production can arise in models such as those with new, light, coloured scalars \cite{PhysRevD.86.095023}, or direct $t\bar{t}hh$ vertices \cite{Grober:2010yv,Contino:2012xk}.

ATLAS has carried out a search in the \bbgg final state~\cite{Aad:2014yja}, setting limits on both resonant (masses between 260\,\GeV and 500\,\GeV) and non-resonant Higgs boson pair production. CMS has searched in the multi-lepton and multi-lepton+photons final-states in the context of 2HDM extensions of the Higgs sector~\cite{Khachatryan:2014jya}. 
CMS has also searched for narrow resonances in the $b\bar{b}b\bar{b}$ channel~\cite{Khachatryan:2015yea}.

Recent phenomenological studies have demonstrated that despite the fully hadronic final state being subject to a large multijet background, searches for new physics in the \pptofourb process have good sensitivity for both resonant~\cite{PhenoBBBB,GouzevitchBBBBPheno} and non-resonant signals~\cite{DeLimaBBBB}. One contributing factor to this sensitivity is the high expected branching ratio for \hbb. The analysis presented in this paper is designed to search for two high-momentum \bbbar systems with masses consistent with $m_{h}$, where each \bbbar system contains two jets identified as containing $b$-hadrons (the jets are ``$b$-tagged''). Compared to a more inclusive $b\bar{b}b\bar{b}$ final-state analysis, this topology has many benefits due to the large required momentum and angular separation between the two \bbbar systems: (i) excellent rejection of all backgrounds; (ii) highly efficient triggering using $b$-tagged multijet triggers; and (iii) negligible combinatorial ambiguity in forming Higgs boson candidates. 

\begin{figure}[!ht]
\begin{center}
\includegraphics[width=0.60\textwidth]{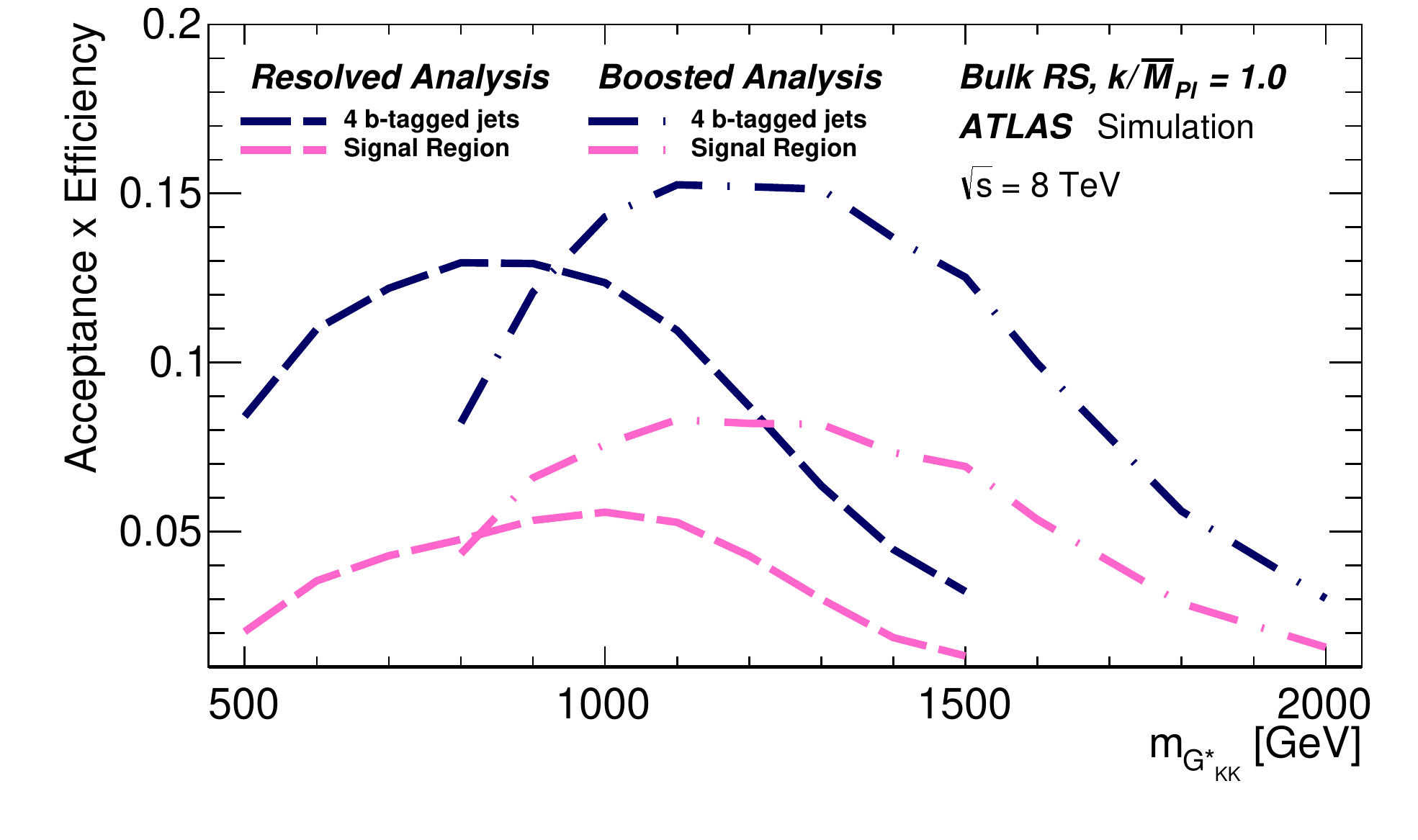}
\caption{Acceptance times reconstruction and selection efficiency as a function of graviton mass for the bulk RS model, for the resolved and boosted analyses. The shapes of the curves are driven by the separation between the $b$-quarks from the Higgs boson decays and the impact on jet clustering. The requirements are defined in Sects.~\ref{sec:resolvedSel} and \ref{sec:BoostedSel}.}
\label{fig:CombinedEff}
\end{center}
\end{figure}

Two Higgs boson reconstruction techniques, which are complementary in their acceptance, are presented. The first---``resolved''---technique reconstructs Higgs boson candidates from pairs of nearby \akt jets~\cite{Cacciari:2008gp} with radius parameter $R = 0.4$, each $b$-tagged with a multivariate $b$-tagging algorithm \cite{ATLAS-CONF-2014-046}. This resolved technique offers good efficiency over a wide range of Higgs boson momenta and so can be used to reconstruct di-Higgs-boson resonances with mass $m_X$ in the range between 500 and 1500~\GeV. The sensitivity is best for this technique in the range $500\leq m_{X} \lesssim 1100$~\GeV. %, which is of most phenomenological interest for the Run-1 dataset, given the integrated luminosity and centre-of-mass energy. 
It can be seen in Fig.~\ref{fig:CombinedEff} however, that the acceptance for four $b$-tagged \akt $R=0.4$ jets decreases for $m_X \gtrsim 1200$~\GeV. This loss of acceptance is due to the increased boost of the Higgs boson, which reduces the average separation between the $b$- and $\bar{b}$-quarks from the Higgs boson decay, %$\Delta R = \sqrt{\left(\Delta\eta\right)^2 + \left(\Delta\phi\right)^2}$, 
$\Delta R = \sqrt{(\Delta\eta)^2 + (\Delta\phi)^2}$, to values below 0.4. This motivates the use of a second---``boosted''---Higgs boson reconstruction technique that maintains acceptance for these higher-mass resonances through the use of jet substructure techniques. The Higgs boson candidate is reconstructed as a single, trimmed~\cite{Krohn2010} \akt $R=1.0$ jet which must be associated with two $b$-tagged \akt $R = 0.3$ track-jets~\cite{ATL-PHYS-PUB-2014-013}.
The use of track-jets with a smaller $R$ parameter allows Higgs bosons with higher transverse momentum (\pt) to be reconstructed.

The analysis is performed with the dataset recorded by ATLAS in 2012 at $\sqrt{s}=8$~\TeV, corresponding to an integrated luminosity of 19.5\,fb$^{-1}$. For the non-resonant search, a counting experiment is performed and the results are interpreted in the context of SM non-resonant Higgs boson pair production. This interpretation is only carried out for the resolved analysis due to its higher sensitivity to such a signal. For the resonant search, a fit to the reconstructed mass spectrum of $hh$ candidates is carried out and the results are interpreted in the context of both bulk RS \Grav (spin-2) and 2HDM CP-even $H$ boson (spin-0) production. In the bulk RS model, the fermion and boson fields can propagate in a warped extra dimension, which has a curvature parameter $k$. This benchmark model is investigated with three coupling constant values, $\kMPl = 1.0,1.5$ and 2.0 ($\bar{M}_{\mathrm{Pl}} = M_{\mathrm{Pl}}/\sqrt{8\pi}$ is the reduced Planck mass), which cover much of the possible parameter space~\cite{Agashe:2007zd}. The 2HDMs considered have CP-conserving scalar potentials (Type-I, Type-II, Lepton-specific and Flipped)~\cite{Branco:2011iw}, in the regime $m_H = m_A = m_{H^{\pm}}$, with the potential parameter that mixes the two Higgs doublets $m_{12}^2~=~m_A^2\tan\beta/(1~+~\tan^2\beta)$. Interpretations are made as a function of $\tan\beta$ and $\cos\left(\beta-\alpha\right)$. The parameter $\tan\beta$ is the ratio of vacuum expectation values of the two Higgs doublets and $\alpha$ is the mixing angle between the two neutral CP-even scalars.

\label{sec:introduction}

\section{The ATLAS detector}
%sec-detector.tex
ATLAS is a multi-purpose particle physics experiment~\cite{ref:AtlasDet} at the LHC. The detector\footnote{ATLAS uses a right-handed coordinate system with its origin at the nominal interaction point (IP) in the centre of the 
detector and the $z$-axis along the beam pipe. The $x$-axis points from the IP to the centre of the LHC ring, and the $y$-axis points 
upwards. Cylindrical coordinates $(r, \phi)$ are used in the transverse plane, $\phi$ is the azimuthal
angle around the beam pipe. The pseudorapidity, $\eta$, is defined in terms of the polar angle $\theta$ as $\eta = -\ln[\tan (\theta/2)]$.}
consists of inner tracking devices surrounded by a superconducting solenoid, electromagnetic
and hadronic calorimeters, and a muon spectrometer. 
The inner tracking system provides charged-particle tracking in the pseudorapidity region $|\eta| < 2.5$ and vertex reconstruction. It consists of a silicon pixel detector, a silicon microstrip
tracker, and a straw-tube transition radiation tracker. 
The system is surrounded by a solenoid that produces a 2\,T axial magnetic field.
The central calorimeter system consists of a liquid-argon electromagnetic sampling calorimeter
with high granularity covering $|\eta| < 3.2$ and a steel/scintillator-tile calorimeter providing 
hadronic energy measurements in the central pseudorapidity range ($|\eta| < 1.7$). 
The endcap and forward regions are instrumented with liquid-argon calorimeters for both 
electromagnetic and hadronic energy measurements up to $|\eta| = 4.9$. The muon spectrometer
is operated in a magnetic field provided by air-core superconducting toroids and includes tracking
chambers for precise muon momentum measurements up to $|\eta| = 2.7$ and trigger chambers covering the range $|\eta| < 2.4$.
A three-level trigger system is used to select interesting events~\cite{Aad:2012xs}.
The Level-1 trigger reduces the event rate to below 75\,kHz using hardware-based trigger
algorithms acting on a subset of detector information. Two software-based trigger levels, referred to collectively as the High-Level Trigger (HLT),
further reduce the event rate to about 400\,Hz using information from the entire detector.

\label{sec:detector}

\section{Data and simulation samples}
%sec-data-mc.tex
The data sample used in this analysis, after applying data quality requirements that include the availability of $b$-jet triggers, corresponds to an integrated luminosity of $19.5\pm0.5$\,fb$^{-1}$. The uncertainty in the integrated luminosity ($2.8\%$) is derived 
following the same methodology as that detailed in Ref.~\cite{Aad:2013ucp}. The data sample is selected by a combination of five triggers requiring multiple jets or $b$-jets, where $b$-jets are identified by a dedicated HLT $b$-tagging algorithm. This combination of triggers is $> 99.5\%$ efficient for signal events passing the offline selection, across the full mass range considered.

Simulated Monte Carlo (MC) event samples are used to model the different signals, as well as the small background contributions from top-quark pair production (\ttbar) and $Z$+jets events.
The dominant multijet background source is estimated directly from data.
Signal samples for both models studied are generated with \madgraph v1.5.1~\cite{Alwall:2011uj,CP3BulkRS}, interfaced to \pythia~v8.175~\cite{Sjostrand:2007gs} for parton showering, hadronization and underlying-event simulation. The Higgs boson mass is set to 125\,\GeV{}. The CTEQ6L1~\cite{Stump:2003yu} leading-order (LO) parton distribution functions (PDFs) are used. 
Table~\ref{tab:CrossSections} provides the calculated cross-sections and widths for
different signal model parameters. The bulk RS model predictions are calculated at leading order using \madgraph. The 2HDM prediction corresponds to the cross-section for gluon-fusion production plus $b$-associated production plus vector-boson-fusion production. The gluon-fusion cross-section is calculated using SusHi~v1.3.0 \cite{Harlander:2012pb,Harlander:2005rq,Harlander:2003ai,Harlander:2002wh,Aglietti:2004nj,Bonciani:2010ms} at next-to-next-to-leading-order (NNLO) accuracy in QCD. For $b$-associated production, an empirical matching of the four- and five-flavour scheme is used~\cite{Harlander:2011aa}. The four-flavour cross-section is calculated at next-to-leading-order (NLO) accuracy in QCD following Refs.~\cite{Dawson:2003kb,Dittmaier:2003ej}, while the five-flavour cross-section is calculated at NNLO in QCD using SusHi. The vector-boson-fusion cross-section at NNLO accuracy in QCD and NLO in electroweak is taken from Ref.~\cite{Heinemeyer:2013tqa} and corrected by a multiplicative factor of $\cos^2\left(\beta - \alpha\right)$~\cite{Branco:2011iw}. The 2HDM branching ratios are calculated using 2HDMC~v1.6.4 \cite{Eriksson:2009ws}.

\begin{table}[tb]
\caption{Computed production cross-sections times branching ratio \sigXfourb and 
 total widths for selected resonance pole mass values. The bulk RS model predictions
 are obtained with \kMPl~=~1.0; both cross-section and width grow as the square of \kMPl. 
 The 2HDM predictions are for a Type-II model with $\cosba=-0.2$ and $\tanb = 1$.}
\begin{center}
\begin{tabular}{ l c c c}
\toprule
 Model & Mass [\GeV{}] & $\sigma\,\times$ B [fb] & $\Gamma$ [\GeV{}] \\ 
\midrule
 Bulk RS 	& 1000 & 1.47  	& 	55 \\
 Bulk RS 	& 1500 & 0.085 	&	90  \\
 2HDM	& 1000 & 5.54 & 130 \\
 2HDM	& 1500 & 0.330 & 332 \\
\bottomrule
\end{tabular}
\label{tab:CrossSections} 
\end{center}
\end{table}
For the \Gtohhb~signal, three sets of MC samples are generated for each of the three coupling values, \kMPl=1.0, 1.5 and 2.0. The variation in these couplings alters both the total \Grav production cross-section and its width. Samples cover a range of \Grav masses $500\leq\mGrav\leq2000$\,\GeV{}. For the \Htohhb~signal, samples are generated covering the range $500\leq m_{H}\leq1500$\,\GeV. Since the width of $H$, $\Gamma_H$, varies non-trivially with the 2HDM parameters, the samples are generated with fixed $\Gamma_H = 1$\,\GeV. In order to make 2HDM interpretations of the results obtained with these fixed width samples, they are corrected to account for the true resonance width at each point in parameter space, as described in Sect.~\ref{sec:results}.

Non-resonant SM \pptofourb events are generated using the exact form factors for the top loop, taken from HPAIR~\cite{PhysRevD.58.115012,Plehn199646}. The cross-section is defined as the inclusive cross-section.
The gluon-fusion production cross-section at NNLO in QCD from Ref.~\cite{PhysRevLett.111.201801} is used, summed with the NLO predictions for vector-boson-fusion, top-pair-associated and vector-boson-associated production from Ref.~\cite{1401.7340}. The resulting cross-section is $\sigma(pp\rightarrow hh\rightarrow b\bar{b}b\bar{b}) = 3.6\pm0.5$\,fb, where the uncertainty term includes the effects of uncertainties in the renormalization and factorization scale, PDFs, $\alpha_S$ and $Br\left(H\rightarrow b\bar{b}\right)$.

The \ttbar~background sample is generated using \powhegbox v1.0~\cite{Nason:2004rx,Frixione:2007vw,Frixione:2007nw,Alioli:2010xd} interfaced to \pythia~v6.426~\cite{Sjostrand:2006za}, with the top mass fixed to 172.5\,\GeV{} and the CT10~\cite{Lai:2010vv} NLO PDF set. The NNLO+NNLL prediction of 253\,pb for the \ttbar~cross-section~\cite{Cacciari:2011hy,Baernreuther:2012ws,Czakon:2012zr,Czakon:2012pz,Czakon:2013goa,Czakon:20142930} is used for normalization.
Single-top background is negligible.

The $Z$+jets sample is generated using \sherpa v1.4.3 \cite{Gleisberg:2008ta} with the CT10 PDF set and the $Z$ boson decaying to two $b$-quarks. 
The $Z$+jets cross-section is taken from an NLO \powhegbox v1.0 \cite{Alioli:2010qp} plus \pythia~v8.165 prediction, which is found to agree well with measurements in the boosted regime~\cite{Zbb}.

The generated MC events are processed with the GEANT4-based~\cite{Geant4} ATLAS detector simulation~\cite{Collaboration:2010wq}. 
Effects of multiple proton--proton interactions 
(pile-up) are simulated using \pythia~v8.1 with the CTEQ6L1 PDF set and the AU2 tune~\cite{MC12AU2}. The simulated events are weighted so that the distribution of the average number of interactions per bunch-crossing matches that in the data. The same reconstruction software is used to process both the data and the simulated samples.

Table \ref{tab:Generators} summarizes the various event generators and PDF sets, as well as
parton shower and hadronization software used for the analyses presented in this paper.

\begin{table}[htb]
\caption{
  Summary of MC event generators, PDF sets, parton shower and hadronization used in the analysis for both signal and background processes. $\dagger$\madgraph was modified \cite{mg5url} to use the exact top-loop form-factors from HPAIR \cite{PhysRevD.58.115012,Plehn199646} for the gluon-fusion production process. 
}
\begin{center}
\small
\begin{tabular}{l l l l}
\toprule
 Model / Process & Generator & PDF & Parton Shower / Hadron. \\ 
\midrule
 Bulk RS: \pptoGtofourb & \madgraph v1.5.1~\cite{Alwall:2011uj,CP3BulkRS} & CTEQ6L1~\cite{Stump:2003yu}  & \pythia~v8.175~\cite{Sjostrand:2007gs} \\
 2HDM: \pptoHtofourb & \madgraph v1.5.1~\cite{Alwall:2011uj} & CTEQ6L1~\cite{Stump:2003yu}  & \pythia~v8.175~\cite{Sjostrand:2007gs} \\
 SM: \pptofourb & \madgraph v1.5.1~\cite{Alwall:2011uj,mg5url}$\dagger$ & CTEQ6L1~\cite{Stump:2003yu} & \pythia~v8.175~\cite{Sjostrand:2007gs} \\
 \ttbar & \powheg v1.0~\cite{Nason:2004rx,Frixione:2007vw} & CT10~\cite{Lai:2010vv} & \pythia~v6.426 \cite{Sjostrand:2006za} \\
 $Z$+jets & \sherpa v1.4.3 \cite{Gleisberg:2008ta} & CT10~\cite{Lai:2010vv} &  \sherpa v1.4.3 \cite{Gleisberg:2008ta} \\
\bottomrule
\end{tabular}
\label{tab:Generators} 
\end{center}
\end{table}

\label{sec:data-mc}

\section{Resolved analysis}
\subsection{Event reconstruction}
\label{sec:resolvedReco}

Jets are reconstructed from topological clusters of calorimeter cell 
energy deposits at the electromagnetic scale~\cite{Aad:2011he} using the \akt jet clustering
algorithm, with a radius parameter of $R = 0.4$. 
The effects of pile-up on jet energies are accounted for by a jet-area-based correction~\cite{Cacciari:2008gn}.
The jets are then calibrated using \pt- and $\eta$-dependent calibration factors based on MC simulations
and the combination of several in situ techniques applied to data~\cite{Aad:2014bia}. Following this, the jets undergo Global Sequential calibration~\cite{Aad:2011he} which reduces flavour-dependent differences in calorimeter response.
If a muon with $\pt > 4$ \GeV{} and $|\eta| < 2.5$ is found
within a cone of $\Delta R=0.4$ around the jet axis, the four-momentum of the muon is added 
to that of the jet (after correcting for the expected energy deposited by the muon in the calorimeter).
Such muons are reconstructed by combining measurements from the inner tracking
and muon spectrometer systems, and are required to satisfy tight muon identification 
quality criteria~\cite{MuonPerf}.
Jets with a significant energy contribution from pile-up interactions~\cite{jvf} are removed using tracking information. For jets with \pt~$<$ 50~\GeV{} and $|\eta| < 2.4$, the \pt~sum of tracks
matched to the jet is calculated and it is required that at least 50\% of this \pt~sum is due to tracks originating from the primary vertex.\footnote{Proton--proton collision vertices are reconstructed requiring that at least five tracks with \pt~$>$ 0.4~\GeV{} are associated
with a given vertex. The primary vertex is defined as the vertex with the highest summed track $\pt^2$.}

Jets with $|\eta| < 2.5$ are $b$-tagged
using the properties of the tracks associated with them, the most important being the impact parameter
(defined as the track's distance of closest approach to the primary vertex in the transverse plane) 
of each track, as well as the presence and properties of displaced vertices. 
The MV1 $b$-tagging algorithm~\cite{ATLAS-CONF-2014-046} used in this analysis combines the above information
using a neural network and is configured to achieve an efficiency of 70\% for tagging 
$b$-jets,\footnote{A jet is labelled as a $b$-jet if a $b$-quark with transverse momentum above 5~\GeV{}~exists within a cone of $\Delta R =0.3$ around the jet axis.}
with a charm-jet rejection of approximately 5 and a light-quark or gluon jet 
rejection of around 140, as determined in an MC sample of \ttbar~events. 
The $b$-tagging efficiency in the simulation is scaled to reproduce the one measured in data \cite{ATLAS-CONF-2014-004}.

\subsection{Selection}
\label{sec:resolvedSel}
\begin{figure}[!ht]
\begin{center}
\subfloat[Bulk RS, \kMPl = 1]{\includegraphics[width=0.5\textwidth]{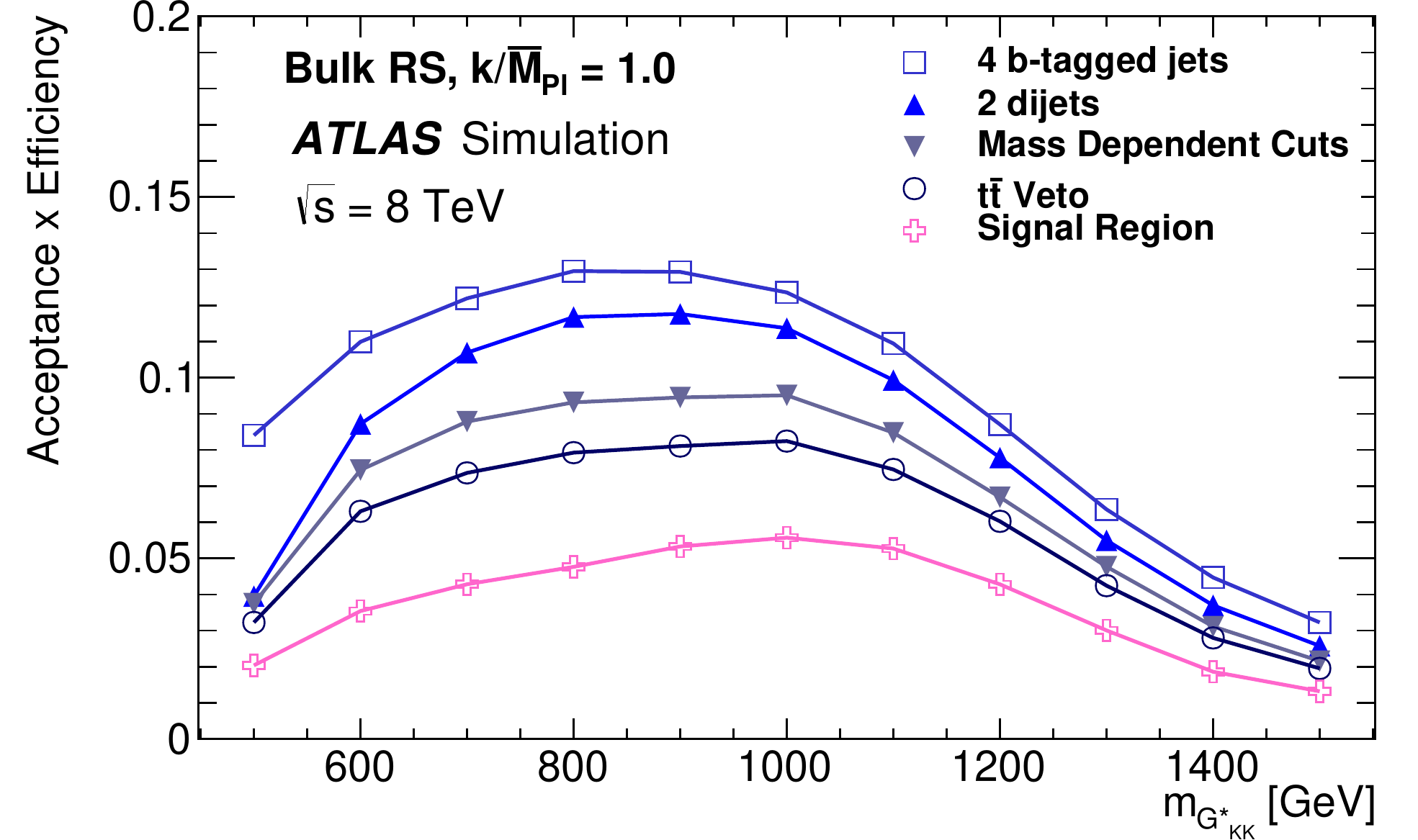}
\label{fig:cutflow10}}
\subfloat[\pptoHtofourb with fixed $\Gamma_H = 1~\GeV{}$]{\includegraphics[width=0.5\textwidth]{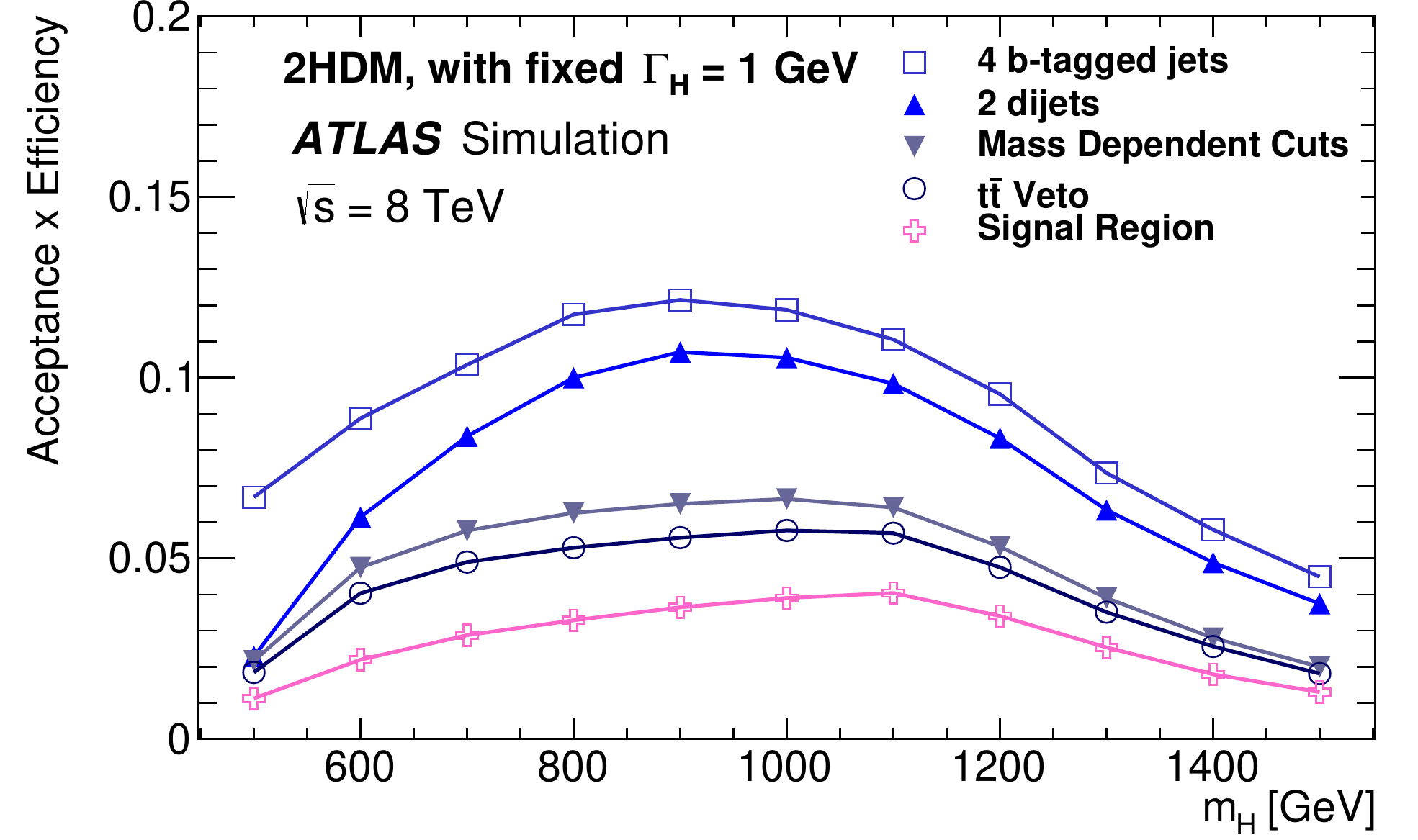}
\label{fig:cutflowPlot2HDM}}
\caption{The selection efficiency as a function of resonance mass at each stage of the event selection for (a) \Gtohhb events and (b) \Htohhb events in the resolved analysis.}
\label{fig:cutflowPlots}
\end{center}
\end{figure}
The combined acceptance times efficiency (\accEff) at different stages of the event selection is shown in Fig.~\ref{fig:cutflowPlots} as a function of resonance mass for the resonant signal models. The event selection begins with the requirement of at least four $b$-tagged jets, each with $\pt > 40$~\GeV{} (shown in Fig.~\ref{fig:cutflowPlots}  as ``4 $b$-tagged jets''). SM non-resonant Higgs boson pair production has a softer Higgs boson \pt spectrum than the $m_X = 500~\GeV{}$ resonances, resulting in a lower $\accEff = 4.9\%$ for this requirement. The four highest-\pt~$b$-tagged jets are then used to form two dijet systems, demanding that the angular distance, $\Delta R$, between the jets in each of the dijets is smaller than 1.5. The transverse momentum of the leading (in \pt) dijet system, \ptlead, is required to be greater than 200~\GeV{}, while the subleading dijet system must have \ptsubl$> 150$~\GeV{}. In the rare case that a jet could be used to create more than one dijet which satisfies the above kinematic requirements, the dijet with the highest mass is chosen. Thus two unique dijet systems, with no jets in common, are selected (shown as ``2 dijets'' in Fig.~\ref{fig:cutflowPlots}). For SM non-resonant Higgs boson pair production, after this requirement, $\accEff = 1.2\%$. The impact of different decay kinematics can be seen by comparing Fig.~\ref{fig:cutflowPlot2HDM} to Fig.~\ref{fig:cutflow10}: the decay of spin-0 $H$ bosons gives a softer Higgs boson \pt spectrum than in the case of the spin-2 \Grav decay (due to the differing angular distributions of hh), resulting in lower acceptance for these kinematic requirements at low resonance mass.

The resolved analysis considers a large range of resonance masses, $500\leq m_X \leq 1500$~\GeV{}. Due to the differing kinematics, the optimal selection for low-mass resonances differs from the optimum for higher masses. To increase the analysis sensitivity, three requirements which vary with the reconstructed resonance mass are used. These selection requirements are optimized simultaneously, by performing a three-dimensional scan of threshold values, using the statistical-only exclusion limit (Sect.~\ref{sec:results}) as the objective function. There are mass-dependent requirements (shown in Fig.~\ref{fig:cutflowPlots}  as ``MDC'') on the minimum \pt of the leading and subleading dijets as well as on the maximum difference in pseudorapidity, $\left|\Delta\eta_{\mathrm{dijets}}\right|$, between them. These requirements are written in terms of the four-jet mass \mfourj\ expressed in \GeV{}: 

\[\pt^{\mathrm{lead}} > \begin{cases}
		400\,\GeV{}& \mathrm{if}\ \mfourj > 910\,\GeV{}, \\
		200\,\GeV{}& \mathrm{if}\ \mfourj < 600\,\GeV{}, \\ 
		0.65\mfourj - 190\,\GeV{}& \mathrm{otherwise},
		\end{cases}
\]
\[\pt^{\mathrm {subl}} > \begin{cases}
	260\,\GeV{}& \mathrm{if}\ \mfourj > 990\,\GeV{}, \\
	150\,\GeV{}& \mathrm{if}\ \mfourj < 520\,\GeV{}, \\ 
	0.235\mfourj + 28\,\GeV{}& \mathrm{otherwise},
	\end{cases}
\]
\[ \dEta < \begin{cases}
	1 & \mathrm{if}\ \mfourj < 820\,\GeV{}, \\
	1.55\times10^{-3}\mfourj - 0.27 & \mathrm{otherwise}.
	\end{cases}
\]

\noindent
The different \mfourj thresholds shown above are chosen to obtain a continuously varying set of requirements.
The requirement on \dEta leads to a lower acceptance for $H$ compared to \Grav for $m_{X} \geq 700$~\GeV{} because of the effect of the boson spin on the angular distribution of its decay products.

After selecting two dijets that satisfy the mass-dependent criteria, \ttbar\ constitutes approximately 10\% of the total background. This \ttbar~background predominantly comprises events where both top quarks decayed hadronically. These hadronic decays lead to three jets for each top quark: one $b$-jet directly from the top decay and two from the decay of the $W$ boson. Since the probability to mis-tag charm-jets is much higher than the probability to mis-tag light-jets, in the majority of cases the dijet is formed from the $b$-jet and a charm-jet from the decay of the $W$ boson. In order to reduce the \ttbar~background, jets not already used in the formation of the two dijets (``extra jets'') in the event are used to reconstruct $W$ boson and top quark candidates by combining them with each of the dijets. These extra jets are required to have $\pt > 30$~\GeV{}, $|\eta| < 2.5$, and $\Delta R<1.5$ relative to the dijet. The $W$ boson candidates are reconstructed by adding the four-momentum of each of the possible extra jets to the four-momentum of the jet in the dijet system with the lowest probability of being a $b$-jet according to the multivariate $b$-tagging algorithm. Top quark candidates are then reconstructed by summing the dijets with each of the extra jets. 
The compatibility with the top quark decay hypothesis is then determined using the variable:

\begin{equation*}
X_{tt}\,=\,\sqrt{\left (\frac{m_{W}\,-\,\tilde{m}_{W}}{\sigma_{m_{W}}}\right )^2 + \left (\frac{m_{t}\,-\,\tilde{m}_{t}}{\sigma_{m_{t}}}\right )^2}\,,
\end{equation*}
\noindent
where $m_{W}$ and $m_t$ are the invariant masses of the $W$ boson and top quark candidates, $\sigma_{m_{W}} = 0.1\, m_{W}$, $\sigma_{m_{t}} = 0.1\, m_{t}$, $\tilde{m}_{W} = 80.4$\,\GeV{}~and $\tilde{m}_{t} = 172.5$\,\GeV{}. The values of $\sigma_{m_{W}}$ and $\sigma_{m_{t}}$ reflect the dijet and three-jet system mass resolutions. 
If either dijet in an event has $X_{tt} < 3.2$ for any possible combination with an extra jet, the event is rejected. This requirement reduces the \ttbar~background by $\sim\,$60\%, whilst retaining $\sim\,$90\% of signal events (shown as ``$t\bar{t}$ Veto'' in Fig.~\ref{fig:cutflowPlots}).

The event selection criteria described above are collectively referred to as the ``4-tag'' selection requirements. These requirements select 1891 events. 

Following the 4-tag selection, a requirement on the leading and subleading dijet masses (\mlead and \msublead, respectively) is used to define the signal
region. The central value of this region corresponds to the median values of the narrowest dijet mass intervals that contain 90\% of the MC signal (these were found to be stable with resonance mass).
The definition of the signal region is

\begin{equation}
X_{hh}\,=\,\sqrt{\left(\frac{\mlead\,-\,124\,\GeV{}}{0.1\,\mlead}\right)^2 + \left(\frac{\msublead\,-\,115\,\GeV{}}{0.1\,\msublead}\right)^2}~, 
\label{eq:Xhh}
\end{equation}
\noindent
where the $0.1\,m_{\mathrm{2j}}$ terms represent the widths of the leading and subleading dijet mass distributions. 
The signal region is defined as $X_{hh} < 1.6$. This corresponds to the kinematical requirements illustrated by the inner region in Fig.~\ref{fig:regions}, albeit with data from the 2-tag sample shown. It is optimized to maximize the expected sensitivity of the search. The acceptance times efficiency of the full selection, including this signal region requirement, is shown in Fig.~\ref{fig:cutflowPlots} as ``Signal Region''. For SM non-resonant Higgs boson pair production, the full selection has an $\accEff = 0.60\%$.

The final step of the Higgs boson pair resonant production search is to perform a fit to the four-jet mass \mfourj in the signal region. The sensitivity of this fit is increased by improving the \mfourj resolution in this region, using the constraint that the two dijet masses should equal the Higgs boson mass, i.e. $\mlead = \msublead = m_h$. To this end, each dijet four-momentum is multiplied by a correction factor $\alpha_{\mathrm{dijet}} = m_h/m_{\mathrm{dijet}}$. This leads to an improvement of $\sim\,$30\% in the signal \mfourj resolution---with a significant reduction of the low-mass tails caused by energy loss---with little impact on the background.

\subsection{Background estimation}
After the 4-tag selection described above, about 95\% of the remaining background in the signal region is expected to originate from multijet events, which are modelled using data. The remaining $\sim\,$5\% of the background is \ttbar~events. The \ttbar~yield is determined from data, while the \mfourj~shape is taken from MC simulation. The $Z$+jets contribution
is $<1$\% of the total background and is modelled using MC simulation. The background from all other sources--including processes featuring Higgs bosons--is negligible.

\subsubsection{Multijet background}
The multijet background is modelled using an independent data sample selected by the same trigger and selection
requirements as described in Sect.~\ref{sec:resolvedSel}, except for the $b$-tagging requirement: only one of the two selected dijets has to be formed from $b$-tagged jets, while the other dijet
can be formed from jets that are not $b$-tagged. This ``2-tag'' selection yields a data sample comprising 485377 events, 98\% of which are multijet events and the remaining 2\% are \ttbar. The predicted contamination by the signal is negligible.

\begin{figure}[!ht]
\begin{center}
\includegraphics[width=0.6\textwidth]{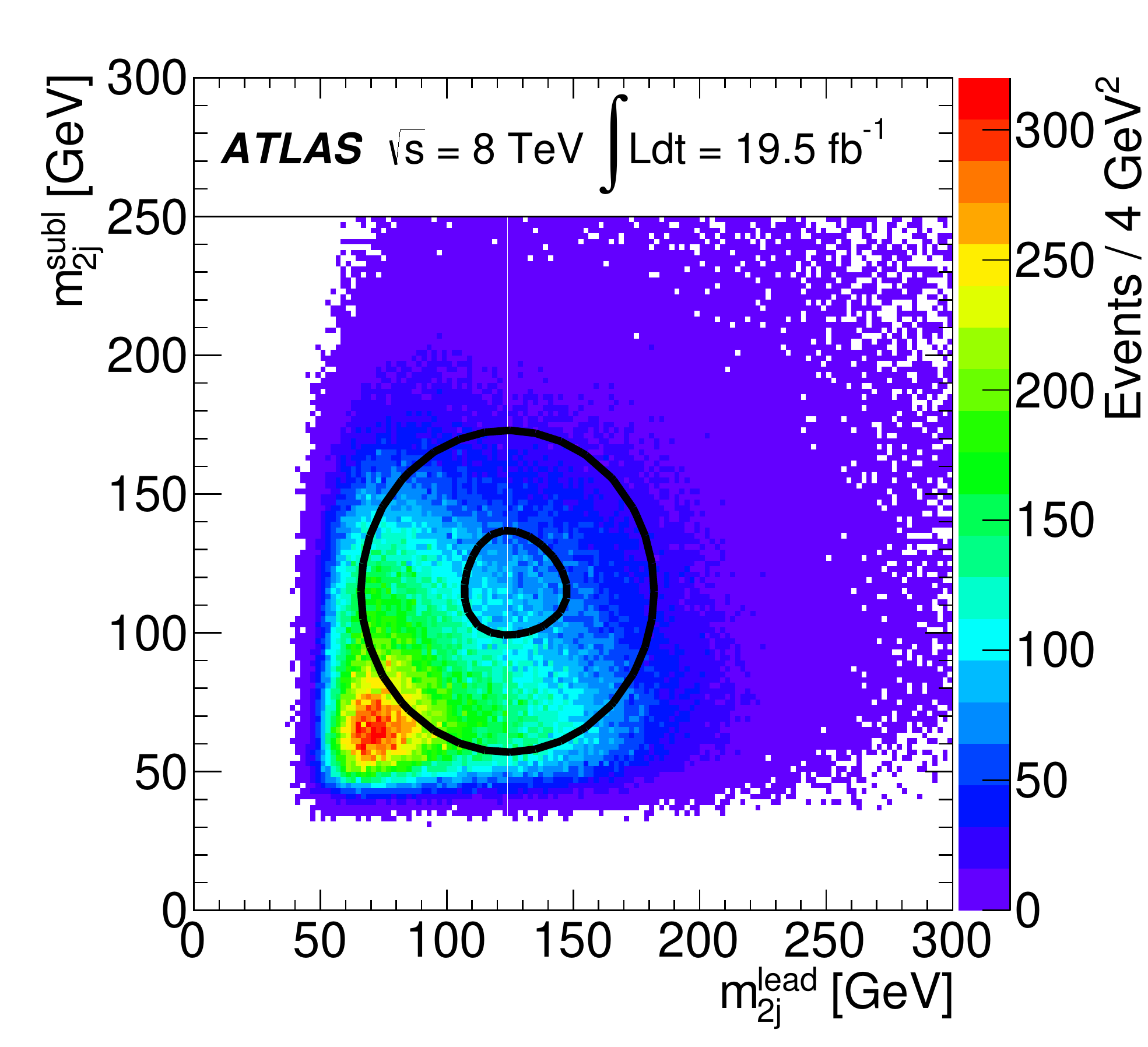}
\label{fig:ControlReg}
\caption{The distribution of the subleading dijet mass, \msublead, vs the leading dijet mass, \mlead, for the 2-tag data sample used to model the multijet background in the resolved analysis. The signal region is the area surrounded by the inner black contour line, centred on $\mlead=124$~\GeV{}, $\msublead=115$~\GeV{}. The control region is the area inside the outer black contour line, excluding the signal region. The sideband region is the area outside the outer contour line.}
\label{fig:regions}
\end{center}
\end{figure}

This 2-tag sample is normalized to the 4-tag sample and its kinematical distributions are corrected for differences introduced by the additional $b$-tagging. These differences arise because the $b$-tagging efficiency as well as the charm- and light-jet rejection vary as a function of jet \pt and $\eta$, the various multijet processes contribute in different fractions, and the fraction of events passed by each trigger path changes. The normalization and kinematic corrections are determined using a signal-free sideband region of the \mlead-\msublead plane, in dedicated samples collected without mass-dependent requirements, which increases the statistical precision of the kinematic corrections. The resulting background model is verified and the associated uncertainties are estimated using a control region. The sideband and control regions are shown in Fig.~\ref{fig:regions}. The sideband region is defined as: $\sqrt{\left(\mlead - 124~\GeV{}\right)^2 + \left(\msublead - 115~\GeV{}\right)^2}~>~58~\mathrm{\GeV{}}$, while the control region is defined as the region between the signal and sideband regions. These definitions are chosen to be orthogonal to the signal region and to give approximately equal event yields in the sideband and control regions. 

The normalization of the multijet background prediction is set by scaling the number of events in each
region of the 2-tag sample by the following factor, \muqcd, calculated in the sideband region:

\begin{equation}
\muqcd = \frac{N^{\mathrm{4-tag}}_{\mathrm{QCD}}}{N^{\mathrm{2-tag}}_{\mathrm{QCD}}} = \frac{N^{\mathrm{4-tag}}_{\mathrm{data}} - N^{\mathrm{4-tag}}_{t\bar{t}} - N^{\mathrm{4-tag}}_{Z}}{N^{\mathrm{2-tag}}_{\mathrm{data}}-  N^{\mathrm{2-tag}}_{t\bar{t}} - N^{\mathrm{2-tag}}_{Z}},
\label{eqn:qcdnorm}
\end{equation}

\noindent
where $N_{\mathrm{data}}^{\mathrm{2-/4-tag}}$ is the number of events observed in the
sideband region in the 2- or 4-tag data sample, respectively. The yields $N_{t\bar{t}}^{\mathrm{2-/4-tag}}$ and
$N_Z^{\mathrm{2-/4-tag}}$ are the estimated number of \ttbar and $Z$+jets events in the
2-/4-tag selected sideband region. The \ttbar~normalization is estimated from data, as described in Sect.~\ref{sec:ttbarBack},
while the $Z$+jets is estimated using MC simulation.

To predict the distributions of the multijet background in each region, the predicted \ttbar and $Z$+jets 2-tag distributions are first subtracted from the 2-tag data sample distribution before the distribution is scaled by \muqcd.

The correction for the kinematic differences between 2-tag and 4-tag samples is performed by reweighting events in the 2-tag sample.
The weights are derived in the sideband region from linear fits to the ratio of the total background model to data for three kinematic distributions which are found to have the largest disagreement between 2-tag and 4-tag events: the leading dijet \pt, the \dr separation between the jets in the subleading dijet, and the \dr separation between the two dijets. The reweighting is done using one-dimensional distributions, but is iterated so that correlations between the three variables are approximately accounted for. Three iterations are found to be sufficient. After the correction process, there is good agreement between the background model and sideband region data in kinematic variables that were not explicitly corrected. 
Systematic uncertainties in the normalization and shape of the multijet background model in the signal region
are assessed using control-region data, as described in Sect.~\ref{sec:ResolvedSyst}.

\subsubsection{\ttbar background}
\label{sec:ttbarBack}
The \ttbar~background is described using a hybrid model: the normalization is derived from data in a \ttbar~control sample, while the shape is taken from MC simulation because there are too few events in the \ttbar~control sample to describe the shape precisely enough.

The \ttbar~control sample is formed from events which pass the 4-tag selection, except for the top veto, which is reversed: if either of the dijets fails the top veto, the event enters the \ttbar~control sample. 
This selection leads to a sample of 41 events within the signal region of the \ttbar~control sample, of which $\sim\,$50\% are \ttbar and $\sim\,$50\% multijet. The multijet background component is estimated using the same methods as used for the nominal selection, but with a wider control region in order to reduce the sideband region \ttbar~fraction. After subtracting the multijet background, the \ttbar~control sample yield is then extrapolated to predict the \ttbar~yield in the nominal signal region, $N_{t\bar{t}}$, using the following equation:

\begin{equation}
  N_{t\bar{t}} = \frac{\epsilon_{t}^{2}}{1-\epsilon_{t}^{2}} \times N_{t\bar{t}}^{\mathrm{CS}} ,
\label{eq:ttbarPred}
\end{equation}
\noindent
where $N_{t\bar{t}}^{\mathrm{CS}}$ is the number of events in the signal region, after subtraction of the multijet background, within the \ttbar~control sample, and $\epsilon_{t}$ is the efficiency for a selected dijet in a \ttbar~event to pass the
top veto. This equation relies on the assumption that the $\epsilon_{t}$ of each dijet in the
event is uncorrelated, an assumption validated in \ttbar~MC simulation. The $\epsilon_{t}$ is measured 
using an independent ``semileptonic \ttbar'' data sample that has a high \ttbar~purity.
Events in this sample are selected by requiring one dijet candidate to pass the nominal selection with $\pt > 150$~\GeV{} and one ``leptonic top-quark'' candidate. 
The leptonic top quark candidate is defined using a reconstructed muon and one $b$-tagged jet.
This $b$-tagged jet is required to be distinct from jets in the dijet
candidate, and the muon is required to have \pt~$>$ 25~\GeV{}, be isolated, and
fall within a cone of radius 1.2 around the $b$-tagged jet. The leptonic
top quark candidate is required to have \pt~$>$ 180~\GeV{}, where the
leptonic top \pt is defined as the vector sum of the $b$-jet
\pt, the muon \pt, and the missing transverse momentum in the event.
The latter is reconstructed from energy deposits in the calorimeter, including corrections
for identified jets, electrons and muons.
The \ttbar~veto efficiency is then measured as the fraction of the reconstructed dijet candidates which passed the \ttbar~veto, yielding $\epsilon_{t} = $ 0.53 $\pm$ 0.03
(stat.) $\pm$ 0.05 (syst.). 
The systematic uncertainty in $\epsilon_{t}$ is assigned to cover potential
differences between $\epsilon_{t}$ as measured in the
semileptonic \ttbar~sample and $\epsilon_{t}$ in the full 4-tag
selection, where the method is applied in \ttbar~MC simulation to evaluate such differences.
The measured $\epsilon_{t}$ agrees well with the corresponding
semileptonic \ttbar~MC prediction of $0.54$.

Equation~(\ref{eq:ttbarPred}) gives a data-driven \ttbar~background prediction of $5.2 \pm 2.6$ events in the signal region after the full selection.
The uncertainty is dominated by the statistical uncertainty in $N_{t\bar{t}}^{\mathrm{CS}}$, with a smaller contribution from the uncertainty in $\epsilon_{t}$.

Due to the limited number of events in the \ttbar~control sample, the \mfourj~shape of the \ttbar~background is modelled using MC simulation. However, despite the use of a large \ttbar~sample, very few events pass the full 4-tag selection. Therefore, the \ttbar~shape is derived from MC simulation using the ``2-tag'' selection, with a systematic uncertainty assigned to cover differences between the 2-tag and 4-tag \mfourj distributions.

\subsection{Systematic uncertainties}
\label{sec:ResolvedSyst}
Two classes of systematic uncertainties are evaluated: those affecting the modelling of the signal and those affecting the background prediction. 

The signal modelling uncertainties comprise: theoretical uncertainties in the acceptance, uncertainties in the jet energy scale (JES) and resolution (JER), and uncertainties in the $b$-tagging efficiency. 

The theoretical uncertainties considered arise from initial- and final-state radiation modelling (ISR and FSR), PDF uncertainties and uncertainty in the LHC beam energy. These are estimated using particle-level samples generated using the same generator configurations as the nominal signal samples but with appropriate variations, assessing the difference in yields after the full analysis selection. The ISR and FSR uncertainty is evaluated by varying the relevant parton shower parameters in \pythia 8. The PDF uncertainty is estimated by taking the maximum difference between the predictions when using MSTW2008nlo \cite{Martin:2009nx}, NNPDF2.3 \cite{Ball:2012cx} and CTEQ6L1. The uncertainty due to the beam energy is determined by varying coherently the energy of each beam by $\pm 26.5\GeV{}$~\cite{Wenninger:1546734} in the simulation. Only FSR has a significant impact on the acceptance, leading to a $\pm1.0\%$ theoretical modelling acceptance uncertainty.

The JES systematic uncertainty is evaluated using 15 separate and 
orthogonal uncertainty components, which allow for the correct treatment of correlations across
the kinematic bins~\cite{Aad:2014bia}. 
The JER uncertainty is evaluated by smearing jet energies according to the 
systematic uncertainties of the resolution measurement performed with data~\cite{Aad:2014bia}. 
For $b$-jets with \pt~$< 300$~\GeV{}~the uncertainty in the $b$-tagging efficiency is
evaluated by 
propagating the systematic uncertainty in the measured tagging efficiency for
$b$-jets~\cite{ATLAS-CONF-2014-004}, which ranges from 2\% to 8\% depending on $b$-jet \pt and $\eta$. However, for the higher resonance masses considered in this analysis, there are a significant number of
events containing at least one $b$-jet with \pt~$> 300$\,\GeV{}. The systematic uncertainties in the tagging
efficiencies of these jets are derived from MC simulation and are larger, reaching 24\% for $\pt > 800$~\GeV{}. 

Systematic uncertainties in the normalization and shape of the multijet background model are assessed in the control region. 
Table \ref{tab:resolvedSBCR} shows the estimated background yields in the control and sideband regions. The control region background prediction agrees with the observed data within the data statistical uncertainty of $\pm3.5\%$. To further test the robustness of the background estimation and the assumptions behind it, predictions are made with different sideband and control region definitions and different $b$-tagging requirements on the 2-tag sample. Redefinitions of the sideband and control region changed the kinematic composition of these regions, enhancing the sideband region in either high mass or low mass dijets and therefore altering the kinematic corrections that are applied. These variations induce a maximum %change of $\pm5.5\%$ in the estimated multijet yield and so the uncertainty in the multijet normalization is set to $\pm6\%$. 
change of $\pm6\%$ in the estimated multijet yield and so the uncertainty is set to this value.
Different $b$-tagging requirements on the $b$-tagged dijet in the 2-tag sample are used in order to change the composition of the sample and to vary the degree of $b$-tagging-related kinematic bias. No additional uncertainty is required.

\begin{table}
\caption{The number of events in data and predicted background events after applying the mass-dependent requirements in preselection and in the sideband and control
  regions for the resolved analysis. The uncertainties are purely statistical. The \ttbar~yield in this table, in contrast to the final result, is estimated using MC simulation.}
\begin{center}
\begin{tabular}{ l c  c  }
\toprule
 Sample & Sideband Region & Control Region  \\ 
\midrule
Multijet      & 907   $\pm$ 3      & 789   $\pm$ 3\\
\ttbar        		 & 25.5    $\pm$ 0.3      & 57.5    $\pm$ 0.4\\
$Z$+jets          	 & 14    $\pm$ 1      & 20    $\pm$ 1\\
\midrule
Total    	 & 947   $\pm$ 3      & 867   $\pm$ 3\\
\midrule
%Data        		 & 952   $(\pm 31)$     & 852   $(\pm 29)$\\	
Data        		 & 952        & 852  \\
\bottomrule
\end{tabular}
\label{tab:resolvedSBCR} 
\end{center}
\end{table}

\begin{figure}[!ht]
\begin{center}
\includegraphics[width=0.6\textwidth]{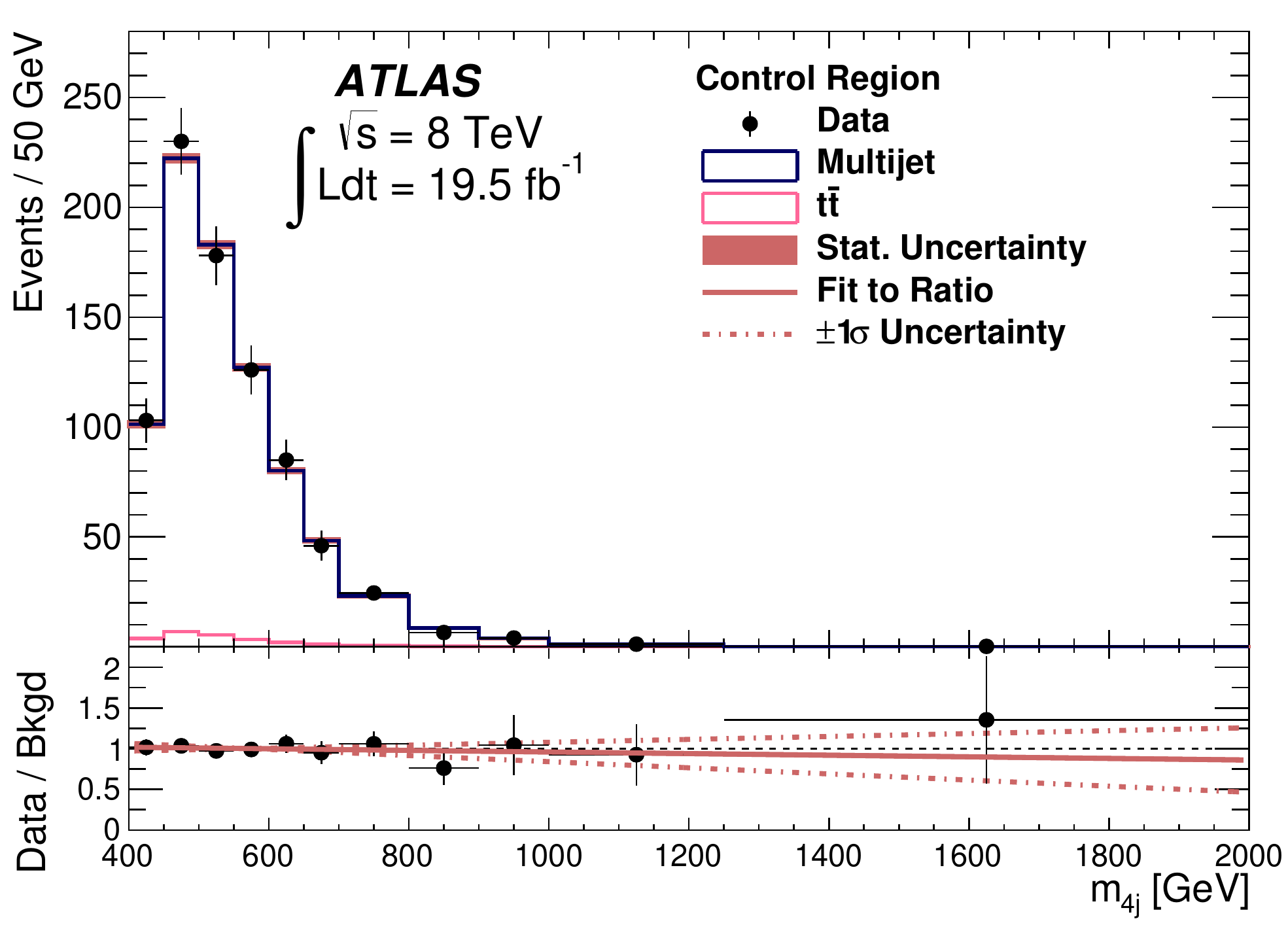}
\caption{The four-jet mass, \mfourj, distribution in the control region (points) for the resolved analysis, compared to the predicted background (histograms). The small filled blocks represent the statistical uncertainty in the total background estimate. The bottom panel shows the first-order polynomial fit to the data-to-background ratio of
  the \mfourj~distribution that is used to fix the multijet shape systematic uncertainty. The dashed lines show the $\pm 1\sigma$ uncertainties in the two fitted parameters.}
\label{fig:ctrlregfit}
\end{center}
\end{figure}

The uncertainty in the description of the multijet \mfourj distribution is determined by comparing the total background prediction to data in the control region, as shown in Fig.~\ref{fig:ctrlregfit}.
Good agreement in the shape is observed and a straight line fit to the ratio of the distributions gives a slope consistent with zero. This fit, along with its uncertainties, is shown in the bottom panel of Fig.~\ref{fig:ctrlregfit}. The uncertainty in the multijet background shape is defined using the uncertainty in the fitted slope.

The uncertainty in the \ttbar~normalization is dominated by the statistical uncertainty of the yield in the \ttbar~control sample, with a subdominant contribution from the uncertainties in the top veto efficiency, $\epsilon_{t}$, leading to a total uncertainty of $\pm50\%$. The uncertainty in the MC-derived \ttbar \mfourj distribution is dominated by the uncertainty associated with using the shape after the 2-tag selection, rather than the 4-tag selection. This uncertainty is assessed by comparing the 2-tag to 4-tag MC predictions in the signal region. A straight line fit to the ratio of the normalized distributions is made and used to define a shape uncertainty in the same way as the multijet background. Due to the large statistical uncertainties of the 4-tag \ttbar~sample, the assigned shape uncertainty is large: $\sim\,$30\% and $\sim\,$100\% in the event yield at \mfourj = 400~\GeV{} and $1500~\GeV{}$, respectively.

\begin{figure}[!ht]
\begin{center}
\includegraphics[width=0.6\textwidth]{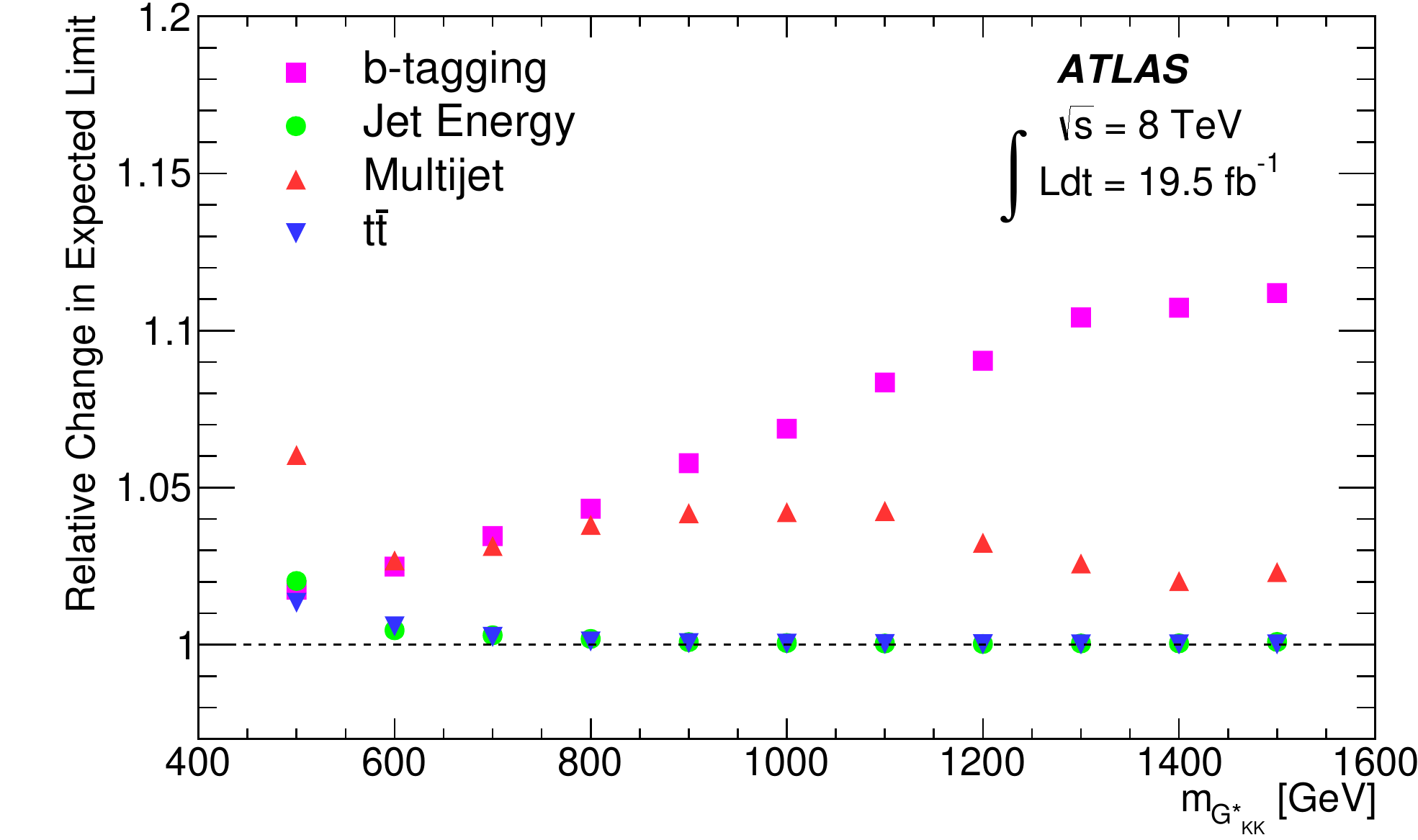}
\caption{The individual impact of the systematic uncertainties considered in the resolved analysis on the expected \sigGfourb 95\% confidence level exclusion limit, as a function of graviton mass. The calculation of the expected limit is described in Sect.~\ref{sec:results}. Only the mass-dependent uncertainties are shown. The impact is the ratio of the limit calculated using all systematic uncertainties sources to the limit calculated using all systematic uncertainty sources excluding those under investigation.}
\label{fig:resolvedSyst}
\end{center}
\end{figure}

Table \ref{tab:resolvedSyst} shows the relative impact of the uncertainties in the event yields. Figure \ref{fig:resolvedSyst} shows the relative impact on the expected limit for \sigGfourb. The calculation of the expected limit is described in Sect.~\ref{sec:results}. It can be seen that for resonance masses below 700~\GeV{}, the effect on the limit is dominated by the multijet description, with a small contribution from the \ttbar background since both backgrounds are predominately at low mass. Above $m_X = 700$~\GeV{}, the uncertainty associated with the modelling of the $b$-tagging efficiency has the largest impact, since the larger high-\pt uncertainties have an increasingly important effect with mass. 

\begin{table}[ht!]
\caption{Summary of systematic uncertainties (expressed in percent) 
 in the total background and signal event yields, in the signal region of the resolved analysis.
 Signal yield uncertainties are provided for non-resonant SM Higgs boson pair production and three resonances with $m=1000~\GeV{}$: a \Grav with \kMPl = 1, another with \kMPl = 2, and $H$ with fixed $\Gamma_H = 1~\GeV{}$.}
\begin{center} 
\begin{tabular}{l@{\hspace{0.18cm}}ccc@{\hspace{0.05cm}}cc}
\toprule
 \vspace{0.09cm}
 Source & Bkgd & SM $hh$ &  \multicolumn{2}{c}{\Grav} & $H$  \\ 
  & &  & \frackMPl = 1 &  \frackMPl = 2 &  \\ 
\midrule
   Luminosity&	--	 & $2.8$ 	& 	$2.8$ 	& $2.8$ 	&$2.8$\\
   JER      	&	--	 & $4.5$ 	& 	$1.1$ 	& 1.1		&$2.0$\\
   JES      	&	--	 & $7$ 	& 	$1.8$	& 1.3		&$3.4$\\
   $b$-tagging&	--	 & $12$ 	& 	$ 22$ 	& 21		&$22$\\
   Theoretical&	--	 & 1.0	&	1.1		&	1.1	&1.1	\\	
   Multijet &	6.0	 & 	-- 		& 	-- 		&	--	&--\\
   \ttbar\  	&	3.0	 & 	-- 		& 	-- 	&	--	&--\\
   \midrule
   Total  	&	6.7	 & $15$ 	& 	$22$ 	& 22		&$23$\\

\bottomrule
\end{tabular}
\label{tab:resolvedSyst}
\end{center}
\end{table}

\subsection{Results of the resolved analysis}
\label{sec:ResolvedResults}
Table~\ref{tab:resolvedResults} shows the predicted number of background events in the signal region, the number of events observed in the data, and the 
predicted yield for two potential signals. The numbers of predicted background events and observed events are in excellent agreement. 
\begin{table}[!ht]
\caption{The number of predicted background events in the $hh$ signal region for the resolved analysis, compared to the data. Uncertainties correspond to the total uncertainties in the predicted event yields. The yield for two potential signals, SM non-resonant Higgs boson pair production and a 500~\GeV{} \Grav in the bulk RS model with
\kMPl = 1 are shown, with the uncertainties taken from Table \ref{tab:resolvedSyst}.}
\begin{center}
\begin{tabular}{ l c  }
\toprule
 Sample & Signal Region Yield \\ 
\midrule
Multijet	& 81.4   $\pm$ 4.9     \\
\ttbar  &   5.2    $\pm$ 2.6   \\
$Z$+jets	& 0.4     $\pm$ 0.2     \\
\midrule
Total 	& 87.0 $\pm$ 5.6       \\
 \midrule
Data 	& 87    \\
\midrule
SM $hh$	& 0.34 $\pm$ 0.05 \\
\Grav$\left(500\,\GeV{}\right)$, \kMPl = 1 & 27 $\pm$ 5.9 \\
\bottomrule
\end{tabular}
\label{tab:resolvedResults} 
\end{center}
\end{table}

\begin{figure}[!ht]
\begin{center}
\includegraphics[width=0.6\textwidth]{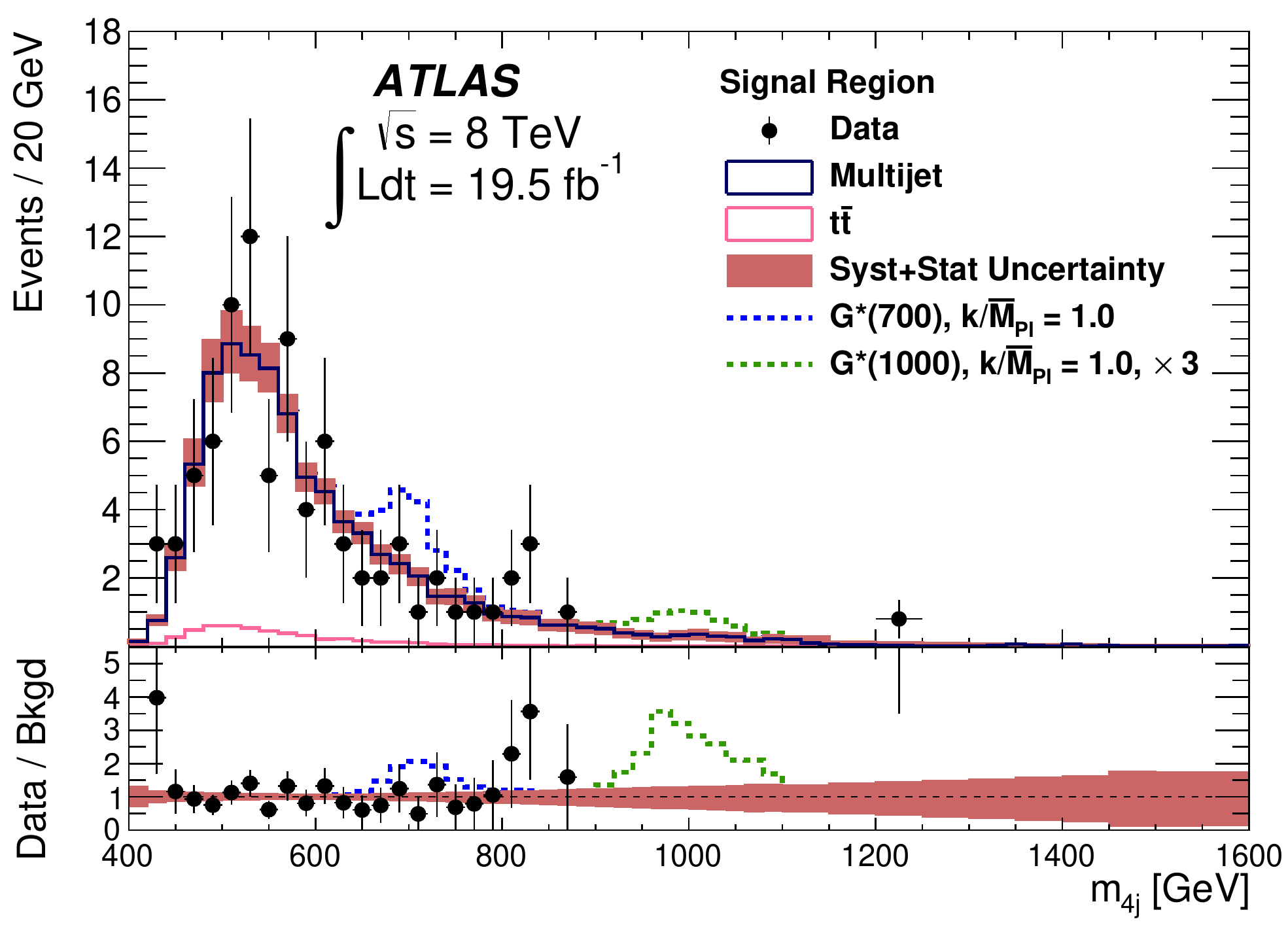}
\caption{Distribution of the four-jet mass, $m_{4j}$, in the signal region of the resolved analysis for data (points) compared to the predicted background (solid histograms). The filled blocks represent the combined statistical and systematic uncertainty in the total background estimate. Two simulated signal \mfourj peaks for the bulk RS model with \kMPl = 1 are shown as dashed lines.}
\label{fig:resolvedHHUnblinded}
\end{center}
\end{figure}

Figure~\ref{fig:resolvedHHUnblinded} shows a comparison of the predicted \mfourj~background distribution to that observed in the data. The predicted background agrees with the observed distributions, 
with no significant deviation. 

\label{sec:resolved}

\section{Boosted analysis}
%
% Event selection

\subsection{Event reconstruction}
The boosted analysis differs from the resolved analysis primarily by
the use of large-radius jets designed to contain the decay products of a single
\hbb decay.
Those large-radius jets, denoted by the subscript J in the remainder of this paper, are reconstructed 
from locally calibrated topological clusters of calorimeter cells~\cite{Aad:2011he} 
using the \akt jet clustering
algorithm with a radius parameter of $R = 1.0$.
To minimize the impact of energy depositions due to pile-up and the underlying event,
the jets are trimmed~\cite{Krohn2010}. This trimming algorithm reconstructs subjets within the large-$R$ jet using the $k_{\mathrm t}$ algorithm with radius parameter $R_{\mathrm{sub}} = 0.3$, then removes any subjet with $\pt$ less than 5\% of the large-$R$ jet \pt.
Further calibration of both the energy and mass scales is applied as a function of \pt and $\eta$ as determined from simulation
and in situ measurements~\cite{Aad:2014bia}.

A novel aspect of the boosted technique presented here is the use of 
track-jets~\cite{ATL-PHYS-PUB-2014-013} to identify the presence of $b$-quarks inside the large-$R$ jet.
Such track-jets are built solely from tracks with $\pt > 0.5$ \GeV{} and $|\eta| < 2.5$, satisfying
a set of hit and impact parameter criteria to make sure that those tracks are consistent with
originating from the primary vertex, thereby reducing the effects of pile-up.
Track jets are reconstructed using the \akt algorithm with $R = 0.3$.
Flavour-tagging of those track-jets proceeds in the same way as for the $R = 0.4$ calorimeter
jets used in the resolved analysis described in the previous section, except for a slightly
looser requirement on the output of the MV1 neural network for a track-jet to be
$b$-tagged. This leads to $b$-jets being $b$-tagged with an efficiency of 74\%, 
with a charm-jet rejection factor of approximately 4 and a light-quark or gluon jet 
rejection factor of around 65, as determined in an MC sample of \ttbar events. 
The $b$-tagging efficiency for track-jets in the MC simulation is adjusted based
on studies of \ttbar events in the data (Sect.~\ref{sec:BoostedSyst}). 

\subsection{Selection}
\label{sec:BoostedSel}
\begin{figure}[!ht]
\begin{center}
\subfloat[Bulk RS, \kMPl = 1]{\includegraphics[width=0.5\textwidth]{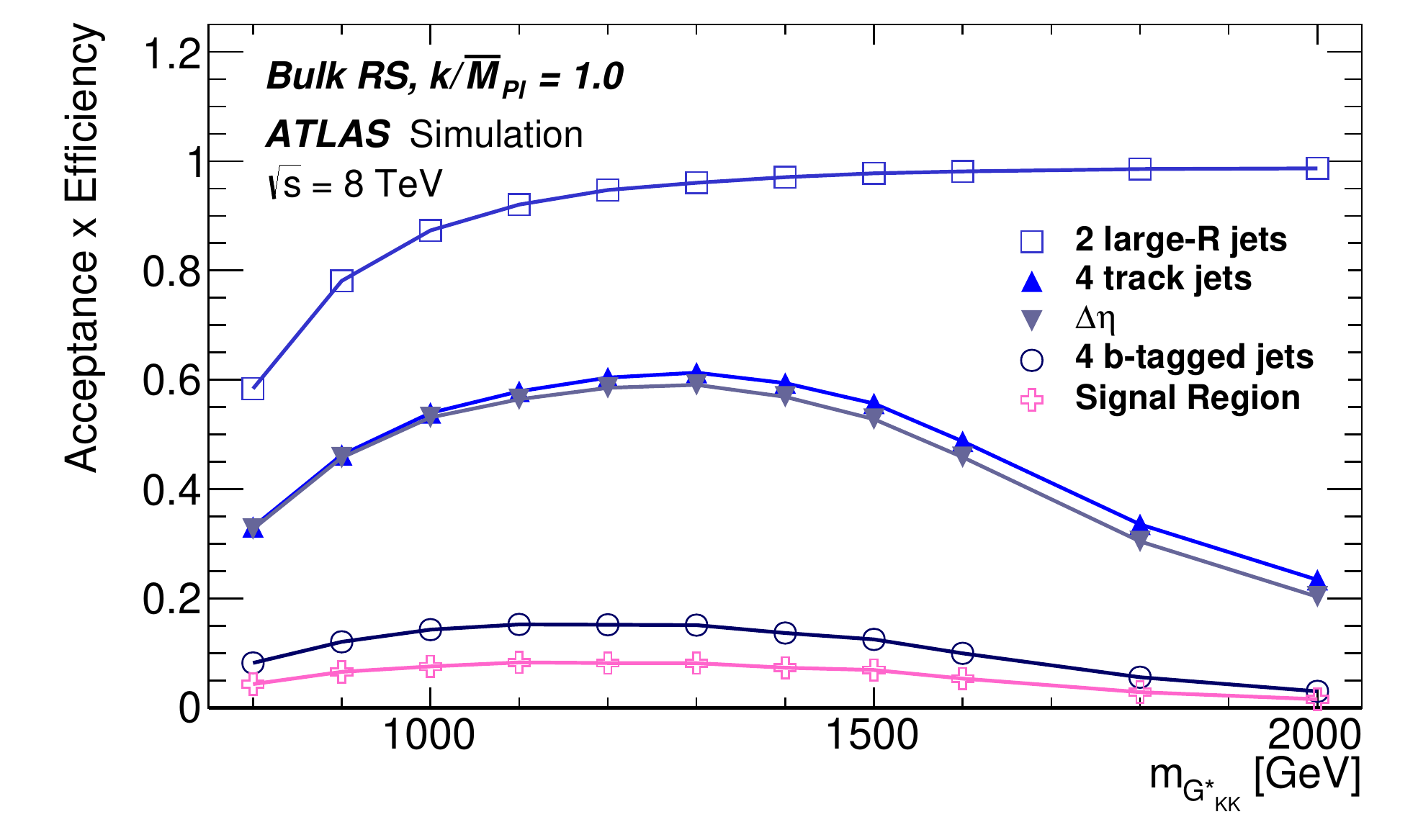}
\label{fig:BoostedCutflowPlotRSG}}
\subfloat[\pptoHtofourb with fixed $\Gamma_H = 1\,\GeV$]{\includegraphics[width=0.5\textwidth]{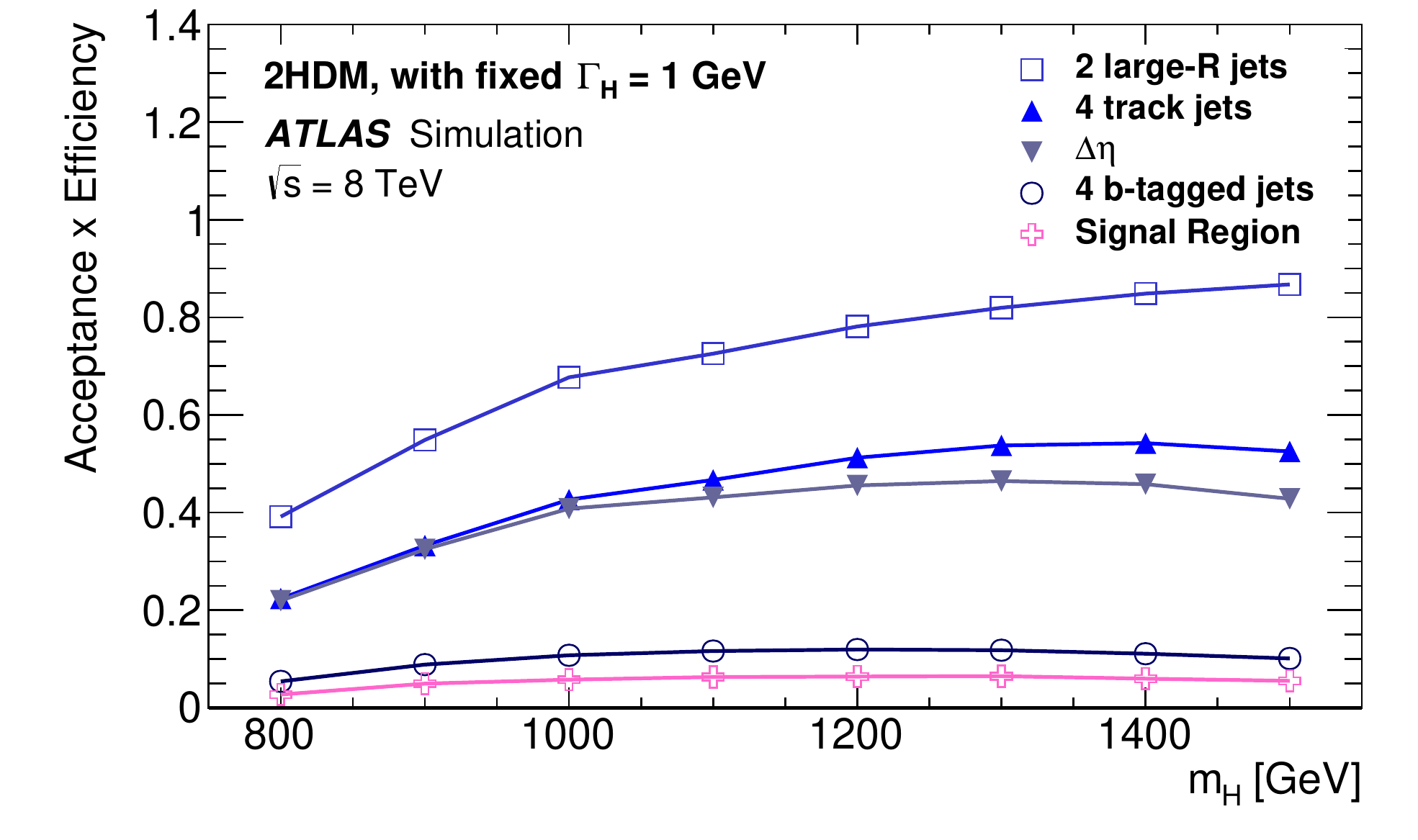}
\label{fig:BoostedCutflowPlot2HDM}}
\caption{The selection efficiency as a function of resonance mass at each stage of the event selection for (a) \Gtohhb events and (b) \Htohhb events
 in the boosted analysis.}
 \label{fig:BoostedCutflowPlot}
\end{center}
\end{figure}

The combined acceptance times efficiency at different stages of the event selection for the boosted analysis is shown in Fig.~\ref{fig:BoostedCutflowPlot}.

Events are required to contain at least two large-$R$ jets with $\pt > 250$ \GeV{} and
$|\eta| < 2.0$. To suppress contamination from \ttbar events, the leading jet is
additionally required to have $\pt > 350$ \GeV{}. This ensures that the top-quark decay
products are typically fully contained in a single large-$R$ jet with mass close
to that of the top quark. These requirements are shown in Fig.~\ref{fig:BoostedCutflowPlot} as ``2 large-R jets''. Only the leading and subleading large-$R$ jets are retained
for further consideration. 

Track jets are associated with large-$R$ jets using
``ghost association'' \cite{Aad:2013gja,Cacciari:2007fd,Cacciari:2008gn}.
Each of the leading and subleading large-$R$ jets must have at least
two track-jets ghost-associated with their respective untrimmed parents, where the
track-jets must have $\pt > 20$ \GeV{} and $|\eta| < 2.5$, as well as be consistent
with originating from the primary vertex of the event (shown in Fig.~\ref{fig:BoostedCutflowPlot} as ``4 track-jets'').
The drop in the $A \times \epsilon$ value at masses above 1500 \GeV{} is due to
the decrease in the angular separation between the two track-jets from the $h \to b\bar{b}$ 
decay to below $\Delta R = 0.3$.

To suppress contamination from multijet events, the two selected large-$R$ jets
in the event are required to have a separation $|\Delta \eta| < 1.7$. This requirement (shown in Fig.~\ref{fig:BoostedCutflowPlot} as ``$\Delta\eta$'') has only
a small impact on the signal acceptance since high-mass resonances tend to produce
jets that are more central than those from multijet background processes.

Selection of $h \to b\bar{b}$ candidates proceeds by requiring that both the leading
and subleading track-jets associated with each of the two
large-$R$ jets satisfy the $b$-tagging selection (shown in Fig.~\ref{fig:BoostedCutflowPlot} as ``4 b-tagged jets'').

A final correction to the large-$R$ jet four-momentum is applied to account for
semileptonic $b$-hadron decays. If a muon passing the requirements outlined in Sect.~\ref{sec:resolvedReco} is ghost-associated with 
any of the selected $b$-tagged track-jets, its four-momentum is added to that of the
large-$R$ jet. If more than one muon is associated with a given track-jet, the muon
closest to the track-jet axis is used.
This correction improves the mass resolution for large-$R$ jets in signal MC simulation, 
especially for the subleading jet.

The last requirement used to select signal event candidates is to require that the large-$R$ jet mass is consistent with the Higgs boson mass. This requirement is defined identically to that for
the resolved analysis in Eq.~(\ref{eq:Xhh}), except for the replacement of
the small-$R$ dijet mass with the large-$R$ jet mass.
The signal region is defined by the requirement $X_{hh} < 1.6$.
This final selection is shown in Fig.~\ref{fig:BoostedCutflowPlot} as ``Signal Region''.

\subsection{Background estimation}
After the 4-tag selection described in Sect.~\ref{sec:BoostedSel}, the background composition is similar to that of the resolved analysis. Multijet events comprise approximately 90\% of the total background and are modelled entirely using data. The remaining $\sim\,$10\% of the background is \ttbar events. The \ttbar yield is determined using data, while the \mtwoJ shape is taken from MC simulation. The $Z$+jets contribution
is $<1$\% of the total background and is modelled using MC simulation. The background from all other sources---including processes featuring Higgs bosons---is negligible.

Estimation and validation of the background described below relies
on two data samples defined as follows.
\begin{itemize}
\item
The ``4-tag'' sample corresponds to the set of events that satisfy all the
requirements detailed in Sect.~\ref{sec:BoostedSel}, except that the
final requirement on the mass of the leading and subleading large-$R$ jets
is not applied.
\item
The ``2+3-tag'' sample is identical to the 4-tag sample except for
having only two or three of the four track-jets $b$-tagged. 
For events with only two $b$-tagged track-jets, both
are required to be associated with the same large-$R$ jet.
\end{itemize}

\begin{figure}[!ht]
\begin{center}
\includegraphics[width=0.6\textwidth]{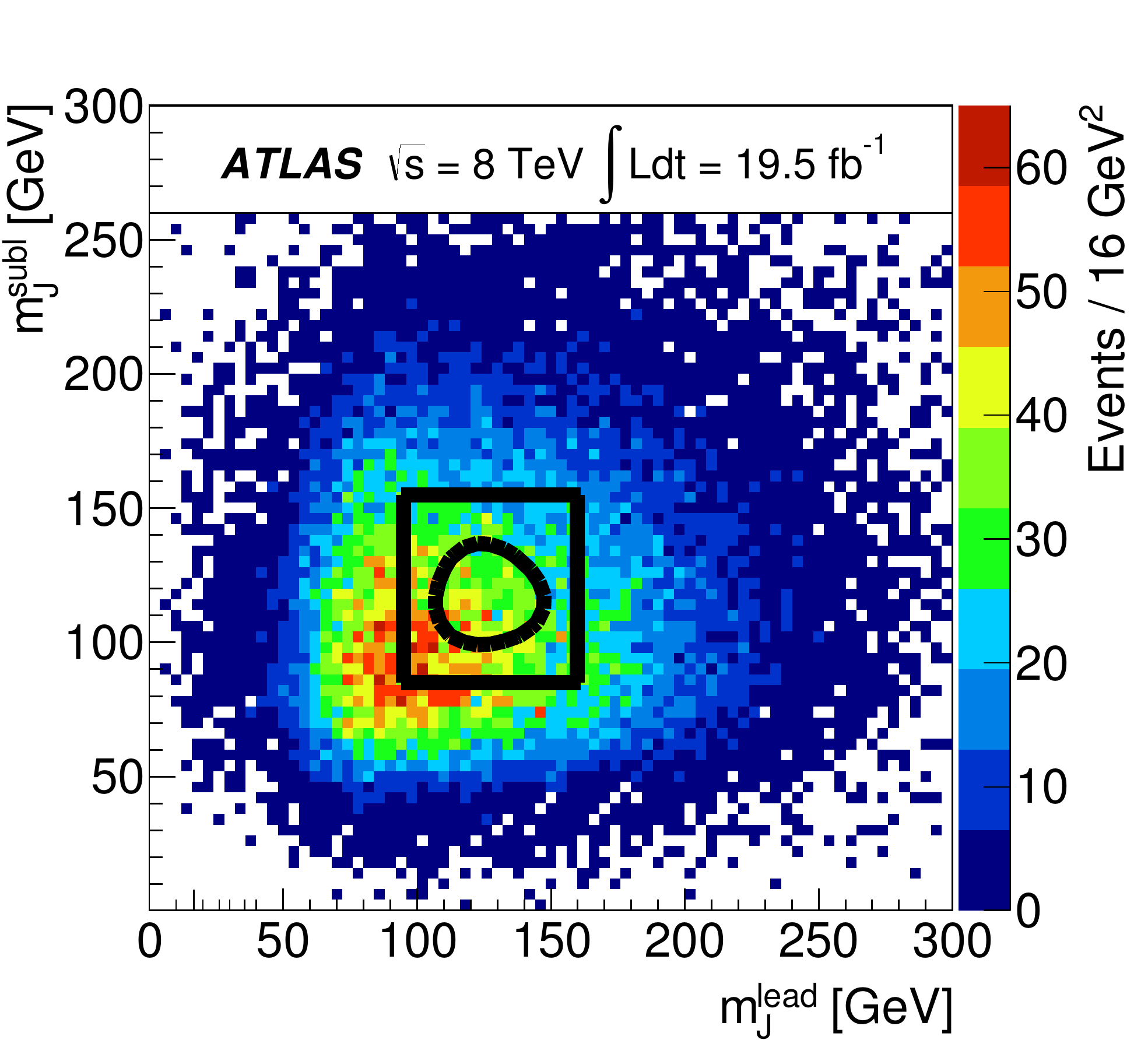}
\caption{The leading--subleading large-$R$ jet mass distribution for the 2-tag and 3-tag data sample
 in the boosted analysis.
 The signal region is the area surrounded by the inner black contour line, centred on
$m_{\mathrm J}^{\mathrm{lead}} = 124$ \GeV{} and $m_{\mathrm J}^{\mathrm{subl}} = 115$ \GeV{}.
The control region is the area inside the outer black contour line, excluding the signal region.
The sideband region is the area outside the outer contour line.}
\label{fig:BoostedRegions}
\end{center}
\end{figure}

Both samples are further subdivided based on the large-$R$ jet masses,
with each subsample having a sideband region to determine the multijet background kinematics
and a control region to validate the background estimate.
The control region is defined by requirements
on the mass of the leading and subleading large-$R$ jets of
$95 < m_{\mathrm J}^{\mathrm{lead}} < 160$ \GeV{} and
$85 < m_{\mathrm J}^{\mathrm{subl}} < 155$ \GeV{} respectively, while excluding the signal region
defined by $X_{hh} < 1.6$.
The sideband region is complementary to the signal and control regions.
Figure~\ref{fig:BoostedRegions} illustrates the sideband and control regions
with data from the 2+3-tag sample.

The choice of control region (and consequently sideband region)
ensures that the multijet background can be estimated
by extrapolation of event yields and kinematic properties from the
2+3-tag sample to the 4-tag sample with a normalization given by the relative event yields 
in the sideband region.
Furthermore, the control region is chosen such that event kinematics in that region
are representative of the kinematics in the signal region.

\begin{figure}[!ht]
\begin{center}
\includegraphics[width=0.6\textwidth]{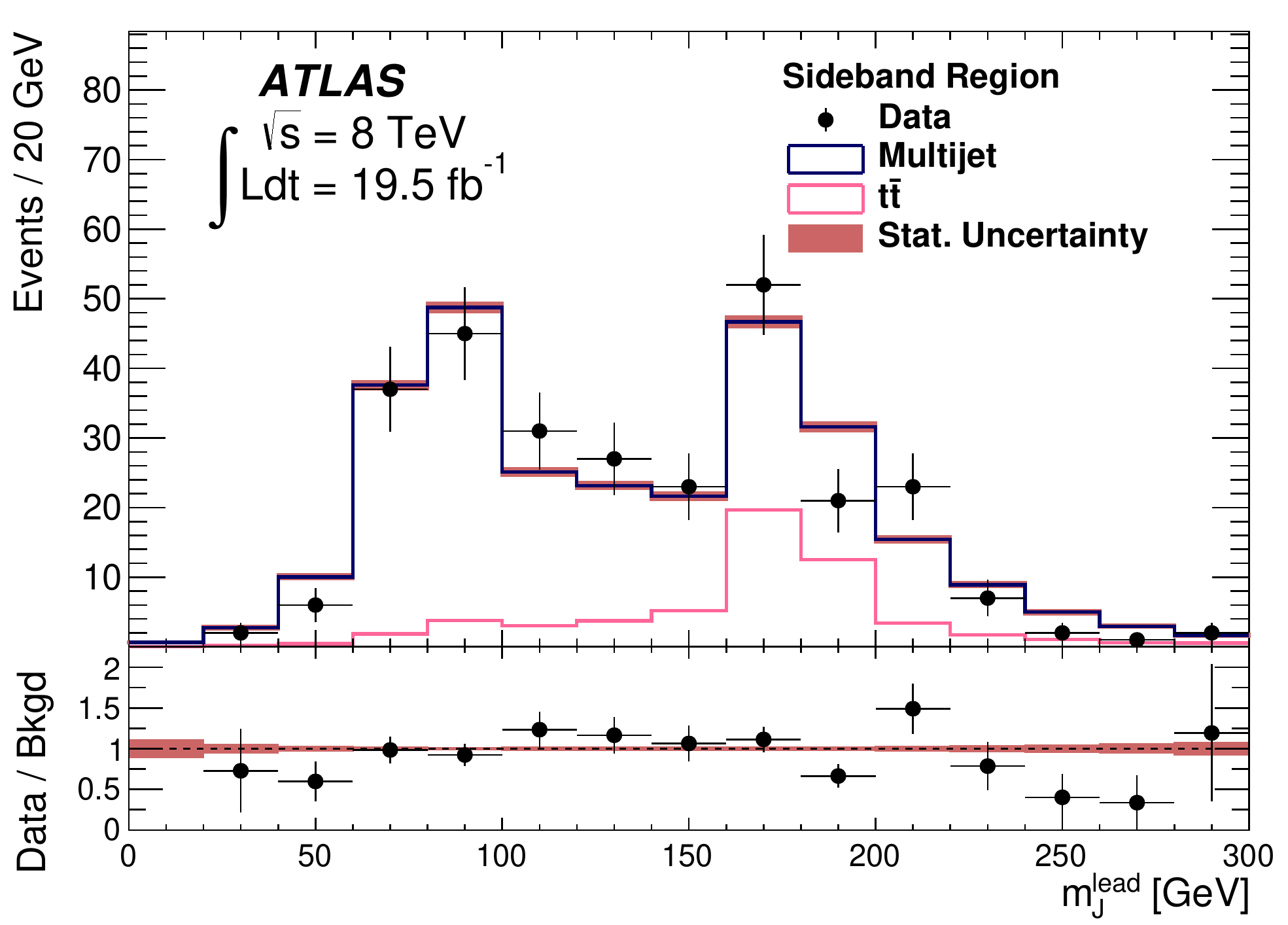}
\caption{Leading large-$R$ jet mass distribution for 4-tag events in the sideband region 
 for data (points) and the two dominant sources of background for the boosted analysis. The normalization for each of those
 two background components is obtained with a fit to the data as described in the text.}
\label{fig:BoostedBackgroundFit}
\end{center}
\end{figure}

The estimated background yield in the 4-tag sample, $N^{\mathrm{4-tag}}_{\mathrm{bkg}}$, is 
computed according to

\begin{equation}
N^{\mathrm{4-tag}}_{\mathrm{bkg}} = \mu_{\mathrm{QCD}}\, N^{\mathrm{2+3-tag}}_{\mathrm{QCD}} + \alpha_{t\bar{t}}\, N^{\mathrm{4-tag}}_{t\bar{t}} + N^{\mathrm{4-tag}}_Z~,
\label{eqn:BoostedBackground}
\end{equation}

\noindent
where $N^{\mathrm{2+3-tag}}_{\mathrm{QCD}}$ is the number of multijet events in the 2+3-tag data sample,
$N^{\mathrm{4-tag}}_{t\bar{t}}$ and $N^{\mathrm{4-tag}}_Z$ are the numbers of events in the 4-tag \ttbar 
and $Z$+jets MC samples.
The parameter $\mu_{\mathrm{QCD}}$ corresponds to the ratio
of multijet event yields in the 4-tag and 2+3-tag data samples, as defined in 
Eq.~(\ref{eqn:qcdnorm}), except for including both 2- and 3-tag events in the denominator.
Finally, the parameter $\alpha_{t\bar{t}}$ is a scale factor designed to adjust the \ttbar
event yield from the MC simulation.
Both $\mu_{\mathrm{QCD}}$ and $\alpha_{t\bar{t}}$ are
extracted from a binned likelihood fit to the leading large-$R$ jet mass distribution obtained in the sideband region of the
4-tag data sample, as depicted in Fig.~\ref{fig:BoostedBackgroundFit}. Due to the large minimum \pt requirement for the leading large-$R$ jet, much of the
\ttbar contribution is concentrated at high mass close to the top-quark mass. 
In this fit, the multijet distribution is extracted from the
2+3-tag data sample, after subtraction of the \ttbar and $Z$+jets contributions
predicted by the MC simulation.
The \ttbar and $Z$+jets distributions in the sideband region of the 4-tag data sample are taken 
from the MC simulation, but the $Z$+jets contribution is very small and its
distribution is added to the multijet distribution for the fit.
The resulting fit values are
$\mu_{\mathrm{QCD}} = 0.0071 \pm 0.0007$ and $\alpha_{t\bar{t}} = 1.44 \pm 0.50$ with
a correlation coefficient of $-0.67$ between these two parameters.

\begin{figure}[!ht]
\begin{center}
\subfloat[Sideband Region]{\includegraphics[width=0.5\textwidth]{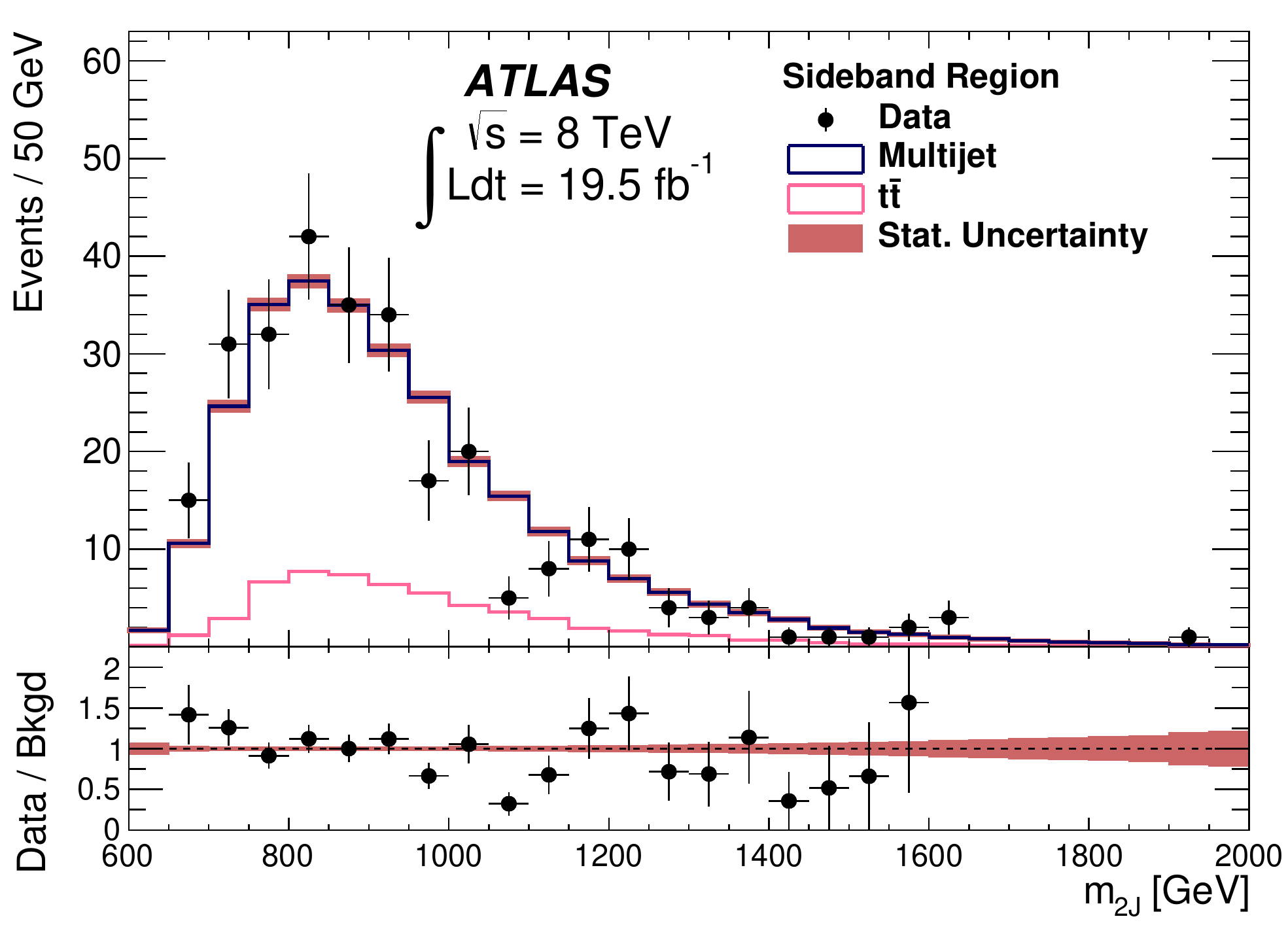}\label{fig:BoostedSidebandDijetMass}}
\subfloat[Control Region]{\includegraphics[width=0.5\textwidth]{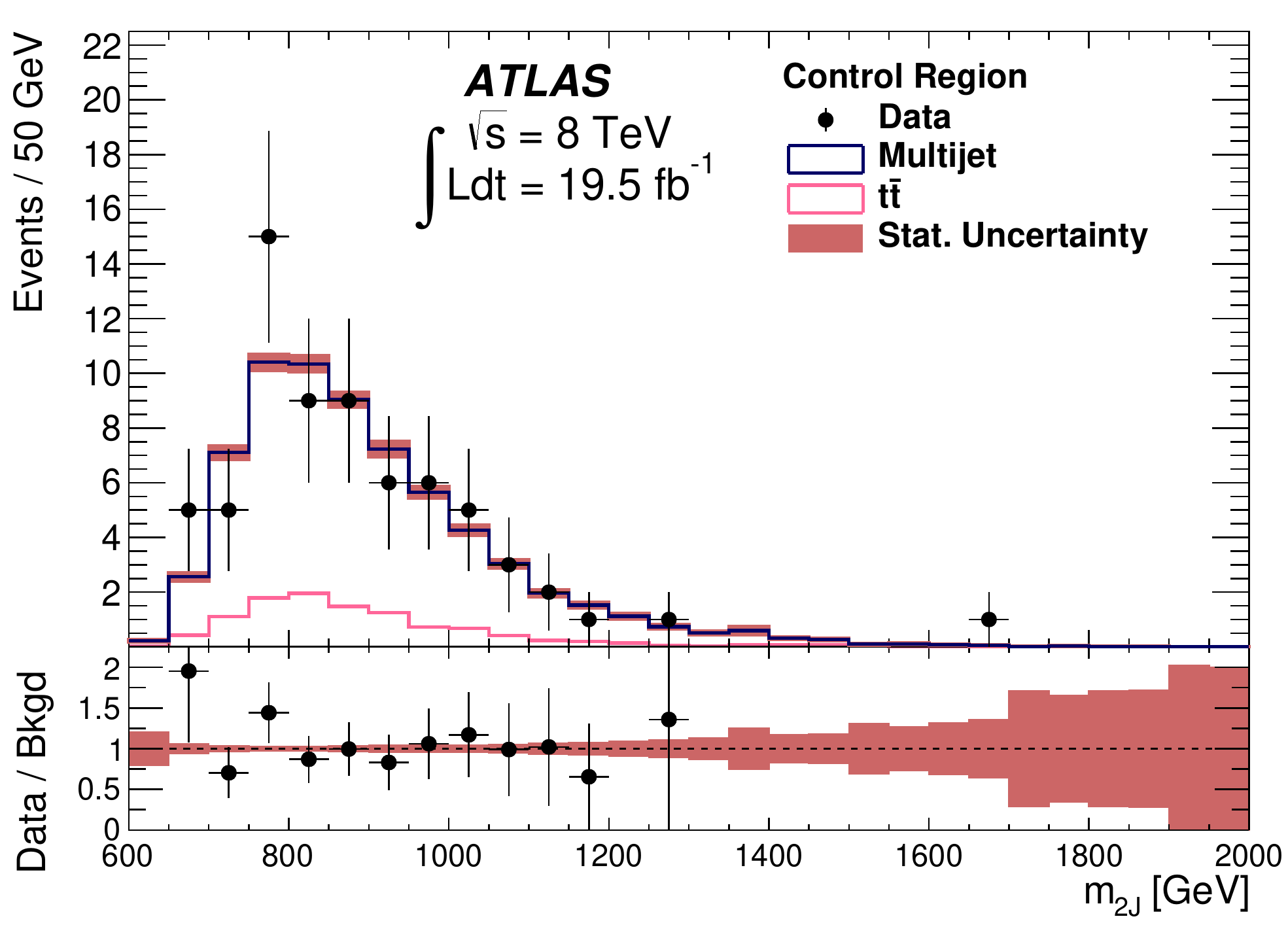}\label{fig:BoostedControlDijetMass}}
\caption{Dijet mass distributions for 4-tag events in the boosted
 analysis. (a) shows the sideband region and (b) the
 control region for data (points) and the expected background (histograms).
 The filled blocks represent the statistical uncertainty in the total background estimate.}
\end{center}
\end{figure}

Figure~\ref{fig:BoostedSidebandDijetMass} shows the dijet mass distribution
for the 4-tag data sample in the sideband region with the background estimated using
the above method. This figure indicates that the 2+3-tag sample provides a valid description
of the background kinematics in the 4-tag sample.
The modelling of the background yield and kinematics is further validated by testing in the control region of the 4-tag data sample.
Good agreement is observed between the data 
and the predicted background in various kinematical distributions for leading and
subleading large-$R$ jets, as well as for the dijet mass, 
as shown in Fig.~\ref{fig:BoostedControlDijetMass}.
The shapes of the \ttbar kinematical distributions in the signal region
are determined from the MC simulation requiring only three $b$-tagged track-jets instead of 
four due to the limited MC sample size. 
The number of \ttbar events is then normalized to the expected yield in the 4-tag sample
times \alphatt. It was checked that this does not
introduce a bias discernible with the statistical precision of the \ttbar MC sample.

\subsection{Systematic uncertainties}
\label{sec:BoostedSyst}
Systematic uncertainties can be grouped into two classes:
those affecting modelling of the signal as extracted from simulation and
those arising from the background estimate.

The signal modelling is affected by two main sources of experimental uncertainty.
One is related to large-$R$ jets and the other is related to the efficiency for
$b$-tagging track-jets.
For large-$R$ jets, the following uncertainties are accounted for:
jet energy scale and resolution, as well as jet mass scale (JMS) and resolution (JMR).
In the kinematic region relevant to this analysis,
the JES uncertainty is below 2\% and that for JMS is $\sim\,$2--5\%.
The ES uncertainty is derived with the $\gamma$--jet balance method for $\pt < 800$ \GeV{}
and the track-jets double-ratio method for $\pt > 800$ \GeV{},
as described in Ref.~\cite{Aad:2013gja}.
The latter method is also used for the derivation of JMS uncertainties in the full \pt range.
An uncertainty of 20\% is applied to account for modelling of the jet energy and mass resolutions.
The magnitude of this resolution uncertainty is estimated from studies of boosted $W$ boson production 
performed using the 2012 data. 
Jet energy and mass uncertainties are treated as uncorrelated in the statistical
analysis.
The uncertainty in modelling the $b$-tagging efficiency for the track-jets used
in this analysis is applied to the signal and $Z$+jets MC samples.
It is extracted as a function of \pt using the tag-and-probe method on a sample
of dilepton events from semileptonic \ttbar decays.
The resulting uncertainty varies between 2\% and 7\%, with the largest value obtained for
track-jets with $\pt > 100$ \GeV{}.
This uncertainty includes the following effects: statistical precision of the calibration
data sample, choice of event generator and parton shower for the simulated \ttbar sample,
initial- and final-state radiation, and flavour composition. For $\pt > 250\,\GeV{}$, the uncertainties must be evaluated using MC simulation due to the small number of data events. Consequently, the uncertainties increase, reaching 14\% for $\pt > 600\,\GeV{}$.
Studies in a \ttbar data sample with a single-lepton+jets final state
indicate that the presence of nearby jets does not have a measurable effect on
the $b$-tagging efficiency and thus no additional uncertainty is required
for nearby jets.

In addition to purely statistical sources of uncertainty, the background estimate
is sensitive to the following other sources.
The multijet background normalization is validated with the observed yield in the control
region and the statistical uncertainty of this test is included as a systematic uncertainty.
The shape of the \ttbar background used in the fit shown in Fig.~\ref{fig:BoostedBackgroundFit}
is varied by extracting the shape from MC samples with zero, one, two or three $b$-tagged track-jets. 
Similarly, the uncertainty in the shape of other \ttbar kinematical distributions is
extracted from those samples.
The uncertainty in the shape of the multijet background extracted from the sideband
region of the 2+3-tag sample
is constrained by the level of agreement between the background prediction and the observed data in
the control region following the procedure described in Sect.~\ref{sec:ResolvedSyst}.
Good agreement is observed between the data and the predicted background in both the sideband
and control regions of the 4-tag sample as shown in Table~\ref{tab:BoostedSBCR}.

\begin{table}
\caption{The number of events in data and predicted background events in the sideband and control
  regions of the 4-tag sample for the boosted analysis. The uncertainties are purely statistical.}
\begin{center}
\begin{tabular}{ l c  c  }
\toprule
 Sample & Sideband Region & Control Region  \\ 
\midrule
Multijet      & 221   $\pm$ 1      & 53.8   $\pm$ 0.6\\
\ttbar        & 52.8  $\pm$ 0.6    &  9.8   $\pm$ 0.3\\
$Z$+jets      & 3.80  $\pm$ 0.26   &  1.57  $\pm$ 0.17\\
\midrule
Total         & 278   $\pm$ 1      & 65.2   $\pm$ 0.7\\
\midrule
%Data          & 281   $(\pm 17)$   & 68     $(\pm 8)$\\	
Data          & 281                & 68     \\	
\bottomrule
\end{tabular}
\label{tab:BoostedSBCR} 
\end{center}
\end{table}

\begin{table}[ht!]
\caption{Summary of systematic uncertainties (expressed in percent) 
 in the total background and signal event yields in the signal region for the boosted analysis.
 Uncertainties are provided for a resonance mass of 1.5 \TeV{} in the context of the
 bulk RS model with \kMPl = 1 or 2, as well as for a Type-II 2HDM with $\Gamma_H = 1$ \GeV{}, $\cosba = -0.2$ and $\tanb = 1$.}
\begin{center} 
\begin{tabular}{  l  c  c  c  c }
\toprule
 Source          & Bkgd    & \multicolumn{2}{c}{\Grav} & $H$ \\
                 &         & \kMPl = 1 & \kMPl = 2  &     \\
\midrule
   Luminosity    &  0.2    &  2.8   &  2.8  & 2.8 \\
   JER           &  0.9    &  0.1   &  0.2  & 0.1 \\
   JES           &  0.4    &  0.1   &  2.5  & 0.1 \\
   JMR           &  4.3    &  13    &  13   & 12  \\
   JMS           &  1.3    &  18    &  17   & 16  \\
   $b$-tagging   &  -      &  21    &  20   & 21  \\
   Theoretical   &  -      &  2.0   &  2.0  & 2.0 \\
   Multijet      &  12     &  -     &  -    &  -  \\
   \ttbar        &  2.5    &  -     &  -    &  -  \\
   Bkgd stat.     &  8.9    &  -     &   -   &  -  \\
   \midrule
   Total         &  15.9   &  33    &  28   &  30 \\
\bottomrule
\end{tabular}
\label{tab:BoostedSyst}
\end{center}
\end{table}

Systematic uncertainties in both the background and signal event yields are summarized
in Table~\ref{tab:BoostedSyst}.
A 2.8\% luminosity uncertainty is applied to the $Z$+jets background and to the signal samples.
The JER/JES/JMR/JMS uncertainties are applied to signal, 
\ttbar\ and $Z$+jets samples. 
The track-jet $b$-tag uncertainty is applied only to the signal samples as the 
normalization and shape differences in the \ttbar\
sample are accounted for through other sources of systematic uncertainty.

Theoretical uncertainties affecting the signal acceptance are also considered, as described
in Sect.~\ref{sec:ResolvedSyst}. These sources do not have significant dependence on the assumed
resonance mass and the largest contribution is found to be due to the ISR modelling.

The uncertainty in the multijet event yield is derived from the difference between the predicted and 
observed multijet yields in the control region.
This source of uncertainty is dominated by the statistical uncertainty in that region.
The \ttbar\ entry in Table~\ref{tab:BoostedSyst} accounts for the shape uncertainty in 
the simulated \ttbar\ leading-jet mass distribution in the 
sideband region used to fit for \muqcd and \alphatt.
This uncertainty is determined by comparing the shape of the 4-tag and 2-tag 
\ttbar\ distributions.
Finally, the ``Bkgd stat'' accounts for the statistical uncertainties in the extraction of
\muqcd and \alphatt. 
Uncertainties in the \mtwoJ shape of the multijet and \ttbar backgrounds are not listed in 
Table~\ref{tab:BoostedSyst},
as they do not affect the event yield, but are accounted for in the statistical analysis.

\begin{figure}[!ht]
\begin{center}
  \includegraphics[width=0.6\textwidth]{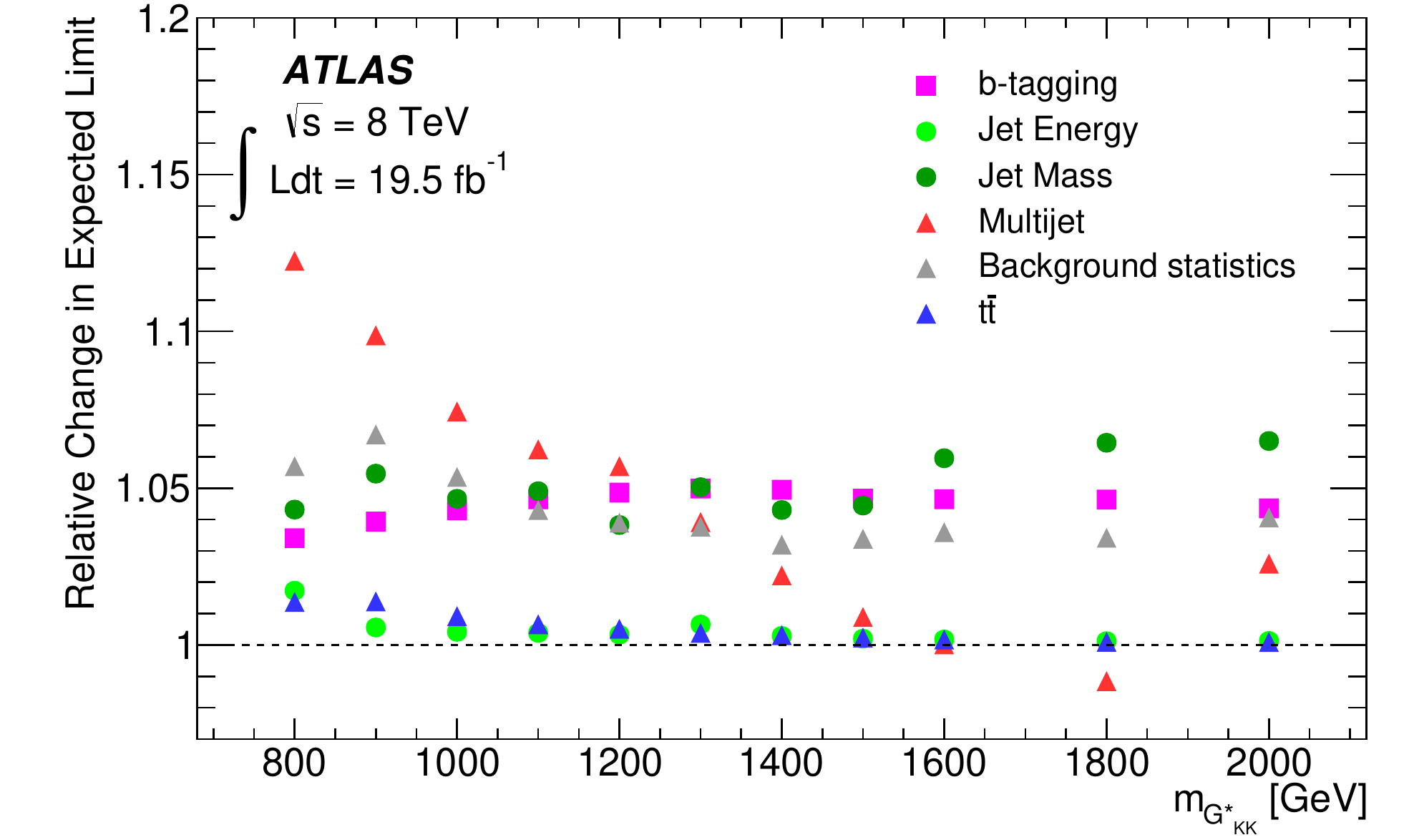}
\caption{The individual relative impact of the systematic uncertainties considered in the
 boosted analysis on the expected \sigGfourb 95\% confidence level exclusion limit, as a function of 
 graviton mass. The calculation of the expected limit is described in Sect.~\ref{sec:results}. Only the mass-dependent uncertainties are shown. The impact is the ratio of the limit calculated using all systematic uncertainties sources to the limit calculated using all systematic uncertainty sources excluding those under investigation.}
\label{fig:BoostedSyst}
\end{center}
\end{figure}

Figure~\ref{fig:BoostedSyst} presents the impact of each source of systematic uncertainty
on the expected cross-section limit for the production of \Grav as a function of resonance mass with the choice \kMPl = 1.
These values are obtained following the statistical analysis described below while neglecting
each source of uncertainty in turn.
The multijet background uncertainty dominates for resonance masses below 1000~\GeV{}, with $b$-tagging,
large-$R$ jet mass and the number of sideband data events for the background estimate becoming
the most important at higher mass.

\subsection{Results of the boosted analysis}

A total of 34 events is observed in the data whereas the background expectation is 
estimated to be $25.7 \pm 4.2$, see Table~\ref{tab:BoostedResults} for a breakdown of 
the various sources of background. The significance of this excess of events in the data is 
evaluated below.

\begin{table}[tb]
\caption{The number of predicted background
  events in the $hh$ signal region, compared to the data for the boosted analysis.
  Errors correspond to the total uncertainties in the predicted event yields.
  The yield for a 1000\,\GeV{} \Grav in 
  the bulk RS model with \kMPl = 1 is also given. }
\begin{center}
\begin{tabular}{ l c }
\toprule
 Sample  & Signal Region Yield \\ 
\midrule
Multijet & 23.5  $\pm$ 4.1 \\ %    2.1 \\
\ttbar 	 &  2.2  $\pm$ 0.9  \\ % 0.7 \\
$Z$+jets &  0.14 $\pm$ 0.06  \\ % 0.05 \\
\midrule
Total 	 & 25.7  $\pm$ 4.2 \\
 \midrule
Data 	 & 34 \\
\midrule
\Grav$\left(1000\,\GeV{}\right)$, \kMPl = 1 & 2.1 $\pm$ 0.6 \\ % 0.1 \\
\bottomrule
\end{tabular}
\label{tab:BoostedResults} 
\end{center}
\end{table}

\begin{figure}[!ht]
\begin{center}
\includegraphics[width=0.6\textwidth]{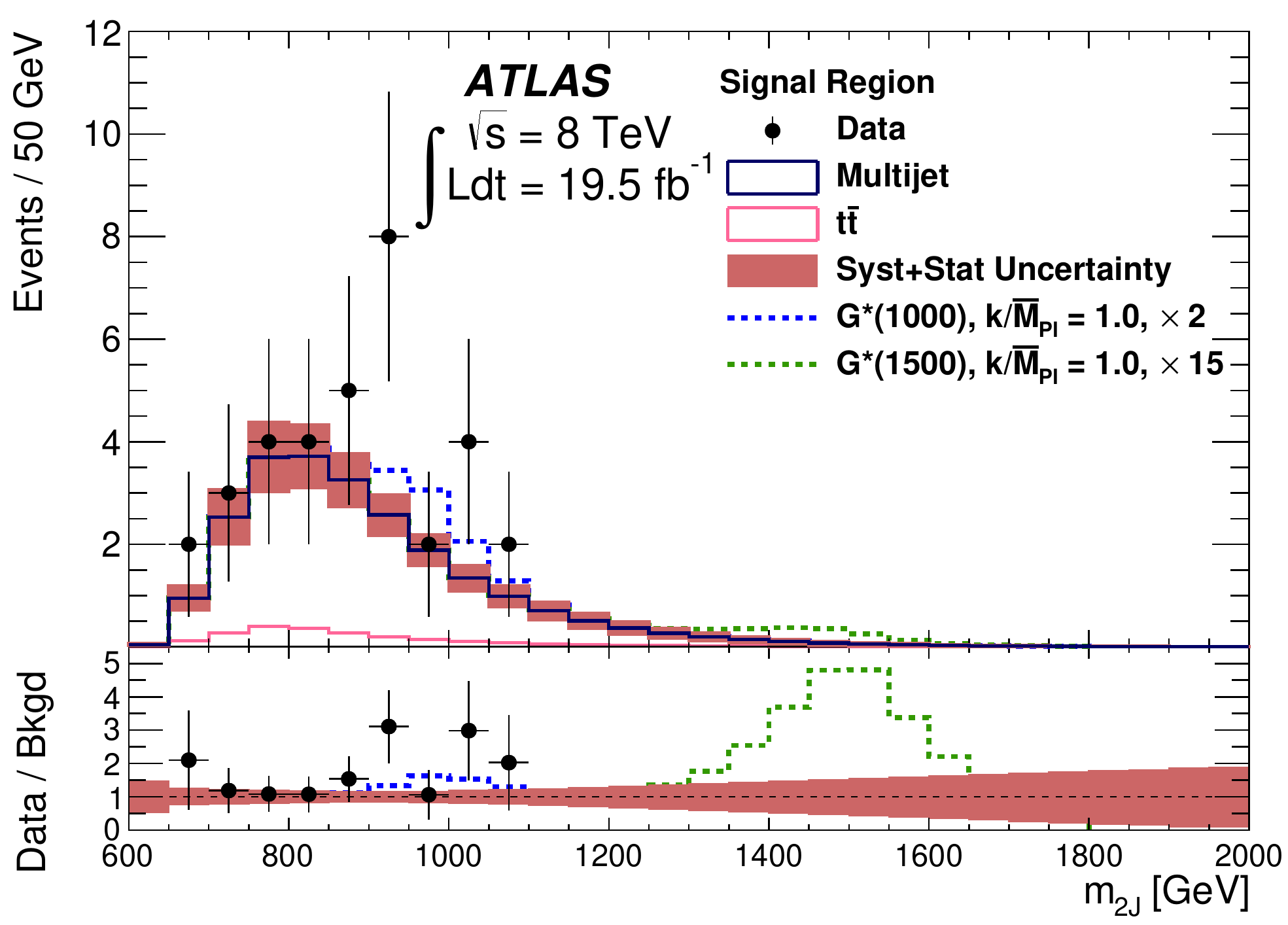}
\caption{Dijet mass distributions for data (points) as well as expected background (solid histograms) in the signal region of the boosted analysis. The filled blocks represent the combined statistical
 and systematic uncertainty in the total background estimate.
 Two simulated \Grav signal peaks predicted by the bulk RS model with \kMPl = 1 are also shown as dashed lines.}
\label{fig:BoostedDijetMassSignalRegion}
\end{center}
\end{figure}

The dijet mass distribution in the signal region is shown in Fig.~\ref{fig:BoostedDijetMassSignalRegion}.
For this distribution and the statistical analysis, the estimated background prediction from
multijet (\ttbar) events is fit to an exponential function at masses above 900 (800) \GeV{} and the
associated uncertainty is propagated to the statistical analysis.

\label{sec:boosted}

\section{Results}
The results from the analyses in Sects. \ref{sec:resolved} and \ref{sec:boosted} are interpreted separately using the statistical procedure described in
Ref.~\cite{Aad:2012tfa} and references therein. Hypothesized values of $\mu$, the global signal strength
factor, are tested with a test statistic based on the
profile likelihood ratio~\cite{Cowan:2010js,Cowan:2013erratum}. In the profile likelihoods, the maximum likelihood values are
obtained with the systematic uncertainties treated as independent,
Gaussian or log-normal constraint terms. 
The statistical analysis described below is performed using data from the signal region solely.
In the case of the search for non-resonant $hh$ production, only the number of events passing
the final selection is used whereas the $m_{\mathrm{4j}}$ or $m_{\mathrm{2J}}$ distributions are used
in the case of the search for $hh$ resonances. 

\subsection{Background-only hypothesis tests}
Tests of the background-only hypothesis
($\mu=0$) are carried out to determine if there are any
statistically significant local excesses in the data. The significance
of an excess is quantified using the local $p_{0}$, the probability that the
background could produce a fluctuation greater than or equal 
to the excess observed in data. A global $p_0$ is also calculated for the most significant discrepancy, using background-only pseudo-experiments to derive a correction for the look-elsewhere effect across the mass range tested.

In the case of the resolved analysis, the largest deviation from the background-only hypothesis is found
to be $2.1\,\sigma$ for a \pptoHtofourb signal with fixed $\Gamma_H = 1\,\GeV{}$ at $m_{\mathrm{4j}} = 1200$~\GeV{}.
This corresponds to a global significance of $0.42\,\sigma$.
The significance of any deviation for a \Grav signal with \kMPl = 1 is very similar,
albeit with slightly smaller local discrepancies as a result of 
the larger signal \mfourj width.

In the case of the boosted analysis, the largest local deviation corresponds to the data excess 
at $\mtwoJ~\sim~900$~\GeV{} apparent in Fig.~\ref{fig:BoostedDijetMassSignalRegion}, with a local 
significance of $2.6\,\sigma$ for \pptoGtofourb with \kMPl = 1. 
The global significance of this deviation corresponds to $0.78\,\sigma$.

Given these low significance values, the results of both analyses are consistent with the background-only hypothesis.
Of the 117 events selected in the data by either the resolved or boosted analysis, only four
events are common to both.

\subsection{Exclusion limits}
The data are used to set upper limits on the cross-sections for the different benchmark signal processes. Exclusion limits are based on the value of the statistic
\cls~\cite{Read:2002hq}, with a value of $\mu$ regarded as excluded at 95\% confidence level (CL) when \cls~is less than 5\%. 

The non-resonant search is performed using the resolved analysis, since it has better sensitivity than the boosted analysis. Using the SM $hh$ non-resonant production as the signal model, the observed 95\% CL upper limit is $\sigma(pp\rightarrow hh\rightarrow b\bar{b}b\bar{b}) = 202$\,fb. This can be compared to the inclusive SM prediction (as defined in Sect. \ref{sec:data-mc}) of $\sigma(pp\hspace{-0.4mm}\rightarrow\hspace{-0.4mm}hh\hspace{-0.4mm}\rightarrow\hspace{-0.4mm}b\bar{b}b\bar{b}) = 3.6 \pm 0.5$\,fb.

\begin{figure}[!ht]
\begin{center}
\subfloat[Resolved Analysis]{\includegraphics[width=0.34\textwidth]{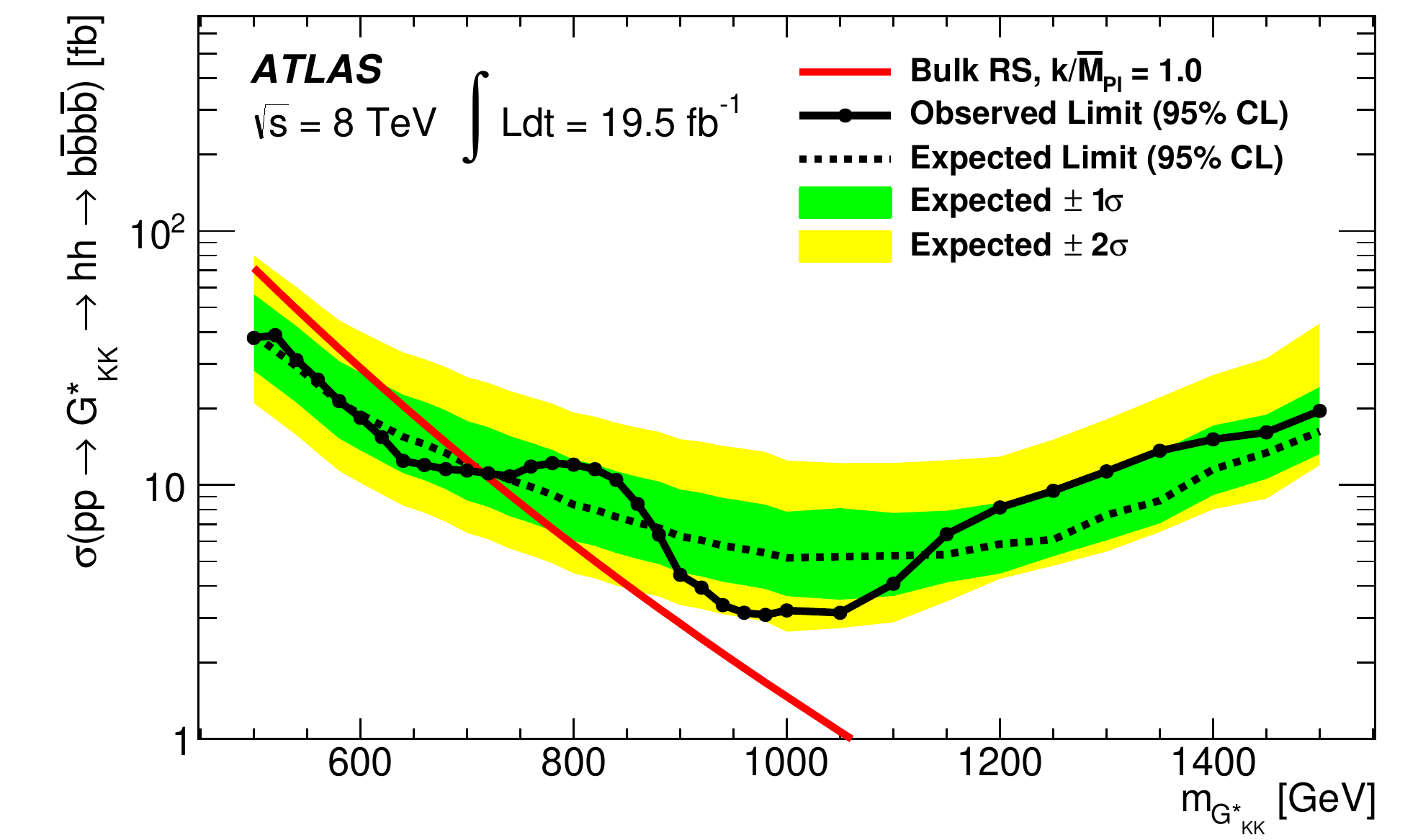}}
\subfloat[Boosted Analysis]{\includegraphics[width=0.34\textwidth]{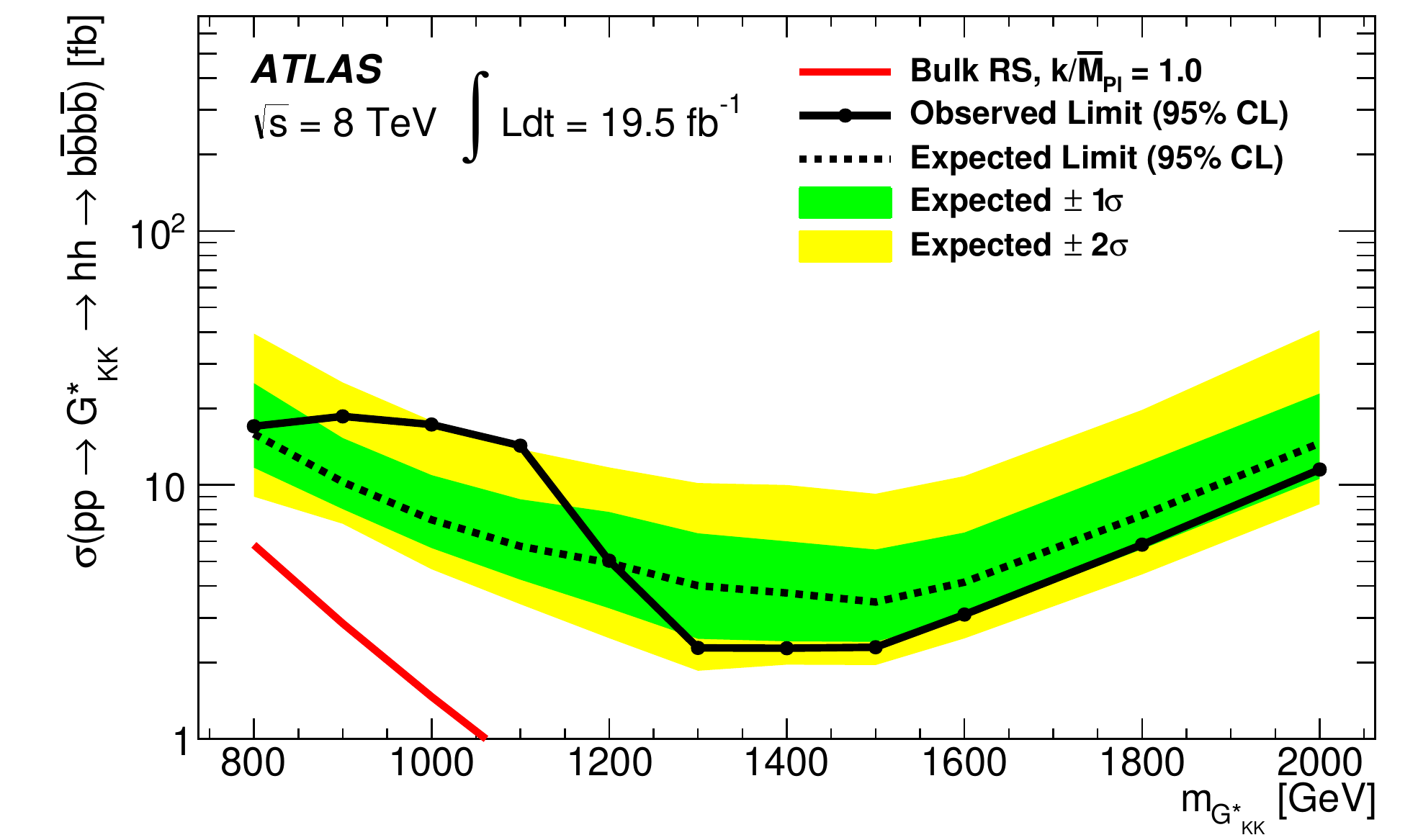}}
\subfloat[Overlay of expected limits]{\includegraphics[width=0.34\textwidth]{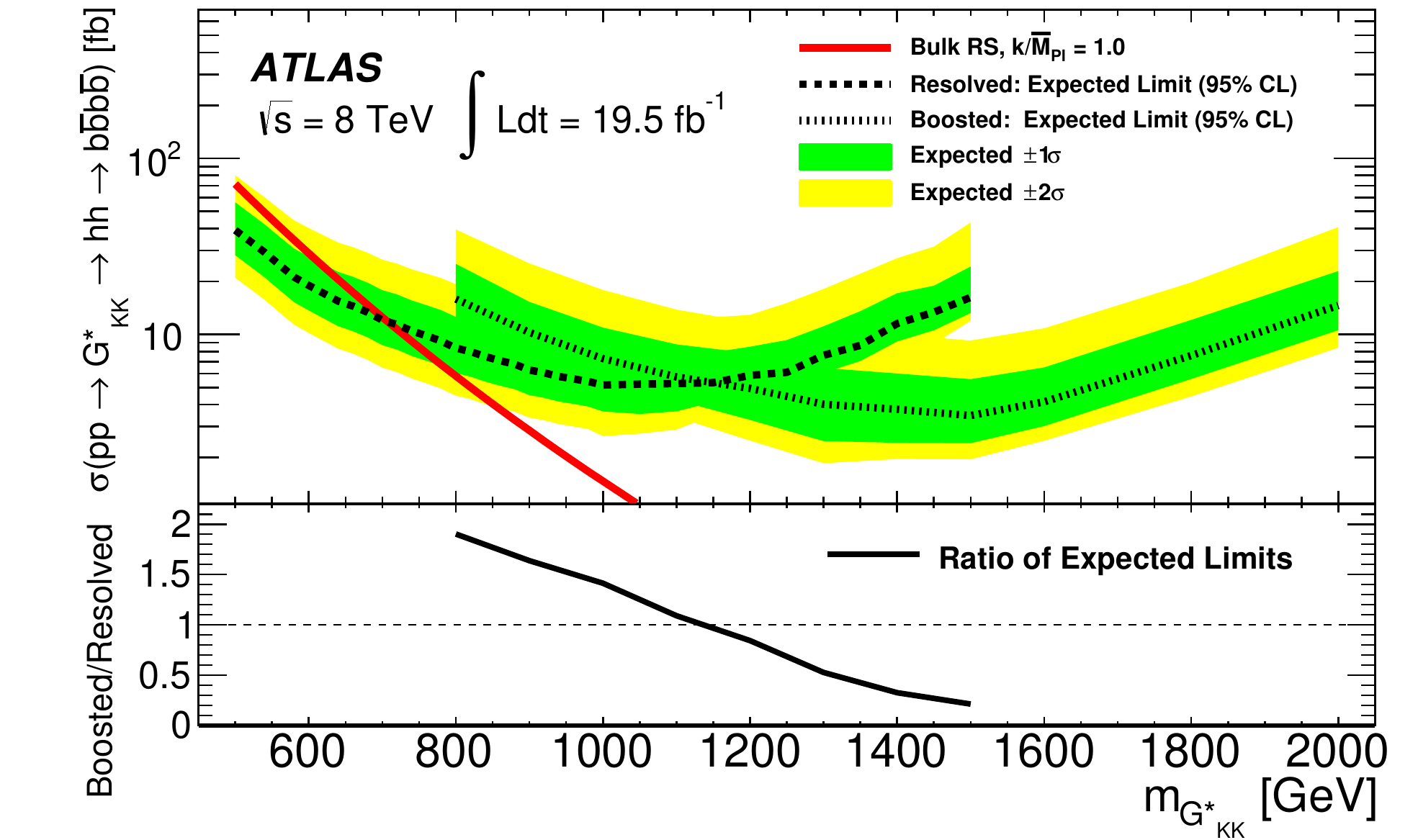}}
\caption{The expected and observed limits for the bulk RS model with $\kMPl=1$ for (a) the resolved analysis and (b) the boosted analysis. The overlay of expected limits is shown in (c), demonstrating that the resolved analysis gives better sensitivity for $\mGrav < 1100\,\GeV{}$, while the boosted analysis is better for $\mGrav > 1100\,\GeV{}$.
The red curves show the predicted cross-section as a function of
resonance mass for the model considered.}
\label{fig:RSGUnblindedLimit}
\end{center}
\end{figure}

For the resonant Higgs boson pair production search, the resolved and boosted analyses offer their best sensitivity in complementary resonance mass regions. Figure~\ref{fig:RSGUnblindedLimit} shows the expected and observed cross-section upper limits from each analysis for \pptoGtofourb within the bulk RS model with $\kMPl=1$. The resolved analysis can be seen to give a more stringent expected exclusion limit for resonance masses up to 1100\,\GeV{}, while the boosted analysis offers better sensitivity beyond that mass. This motivates a simple combination of the separate exclusion limits from the resolved and boosted analyses. For each of the signal models, the limit for each mass point is taken from the analysis which offers the most stringent expected exclusion.

\begin{figure}[!ht]
\begin{center}
\subfloat[Bulk RS, \kMPl = 1]{\includegraphics[width=0.34\textwidth]{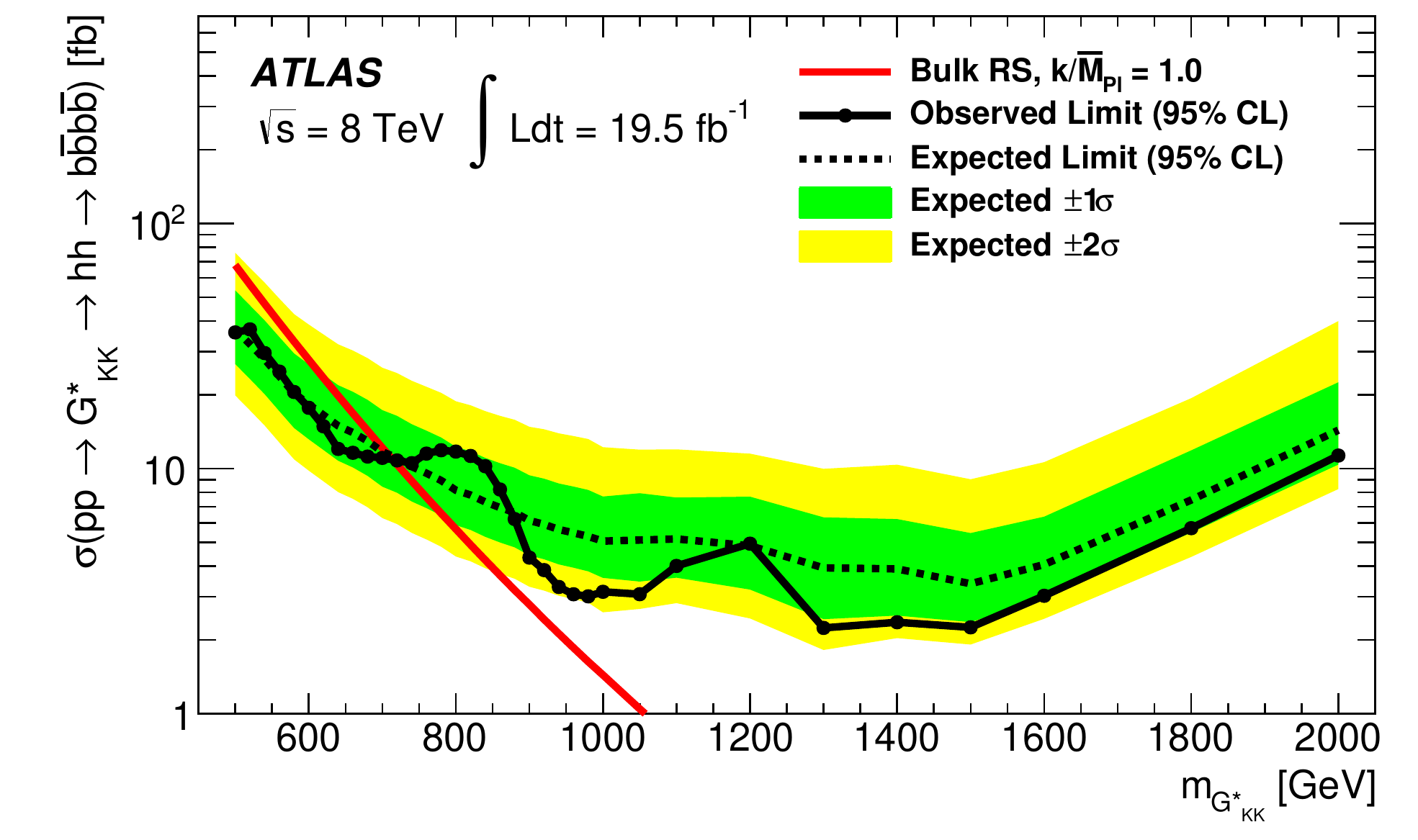}}
\subfloat[Bulk RS, \kMPl = 2]{\includegraphics[width=0.34\textwidth]{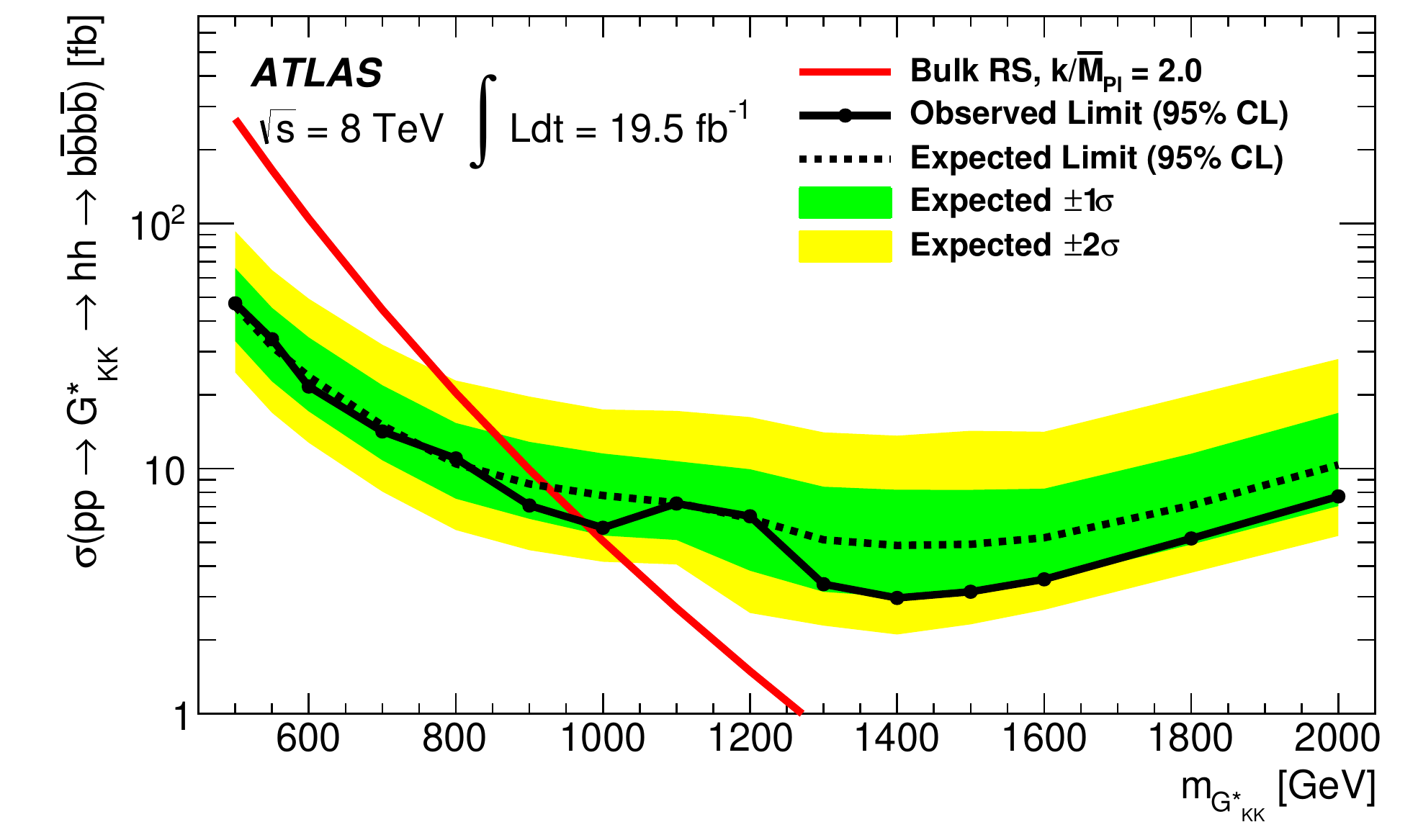}}
\subfloat[\pptoHtofourb with fixed $\Gamma_H = 1\,\GeV{}$]{\includegraphics[width=0.34\textwidth]{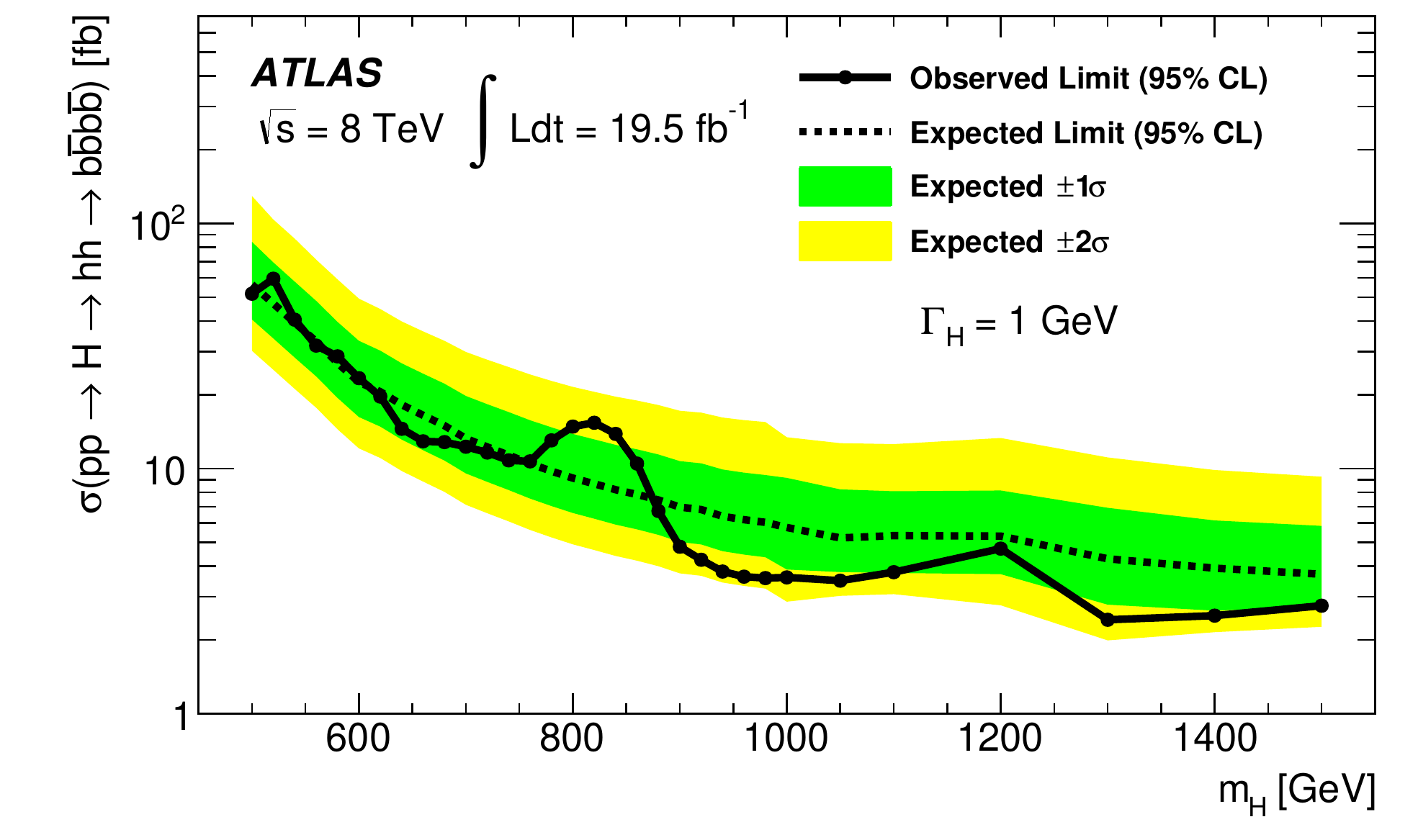}}
\caption{The combined expected and observed limit for \pptoGtofourb in the bulk RS model with (a) \kMPl = 1 and (b) \kMPl = 2, as well as (c) \pptoHtofourb with fixed $\Gamma_H = 1\,\GeV{}$. 
The red curves show the predicted cross-sections as a function of
resonance mass for the models considered.}
\label{fig:CombinedLimits}
\end{center}
\end{figure}

Figure~\ref{fig:CombinedLimits} shows the combined 95\% CL upper limits for three signal models: \pptoGtofourb within the bulk RS model with \kMPl = 1 and 2, and the \pptoHtofourb with a fixed $\Gamma_H = 1\,\GeV{}$. The most stringent limits of $\sigXfourb\sim3\,$fb are set in the range $900 < m_X < 1600\,\GeV{}$, where there is little expected background and either the resolved or boosted analysis provides good signal acceptance. The excluded mass ranges for the bulk RS KK graviton are shown in Table \ref{tab:resolvedGravExcl}. 

\begin{table}[htb]
\caption{Range of KK graviton masses excluded at 95\% confidence level for $\kMPl = 1.0, 1.5$ and 2.0.}
\newcommand\T{\rule{0pt}{2.6ex}}
\newcommand\B{\rule[-1.2ex]{0pt}{0pt}}
\begin{center}
\begin{tabular}{lc}
\hline
\kMPl \T \B & 95\% CL Excluded \Grav Mass Range [\GeV{}] \\
\hline
1.0 \T & $500 - 720$ \\
1.5    & $500 - 800$  and $870 - 910$ \\
2.0 \B & $500 - 990$ \\
\hline
\end{tabular}
\label{tab:resolvedGravExcl}
\end{center}
\end{table}

The excluded mass range for the 2HDM is parameter dependent, principally because the production cross-section varies, but also because the exclusion limit depends on the parameter-dependent $H$ boson width, $\Gamma_H$. The theoretical cross-section used to determine the 95\% CL excluded regions is the sum of the cross-sections of gluon-fusion production, vector-boson-fusion production and $b$-associated production.

\begin{figure}[!ht]
\begin{center}
\subfloat[Type-I 2HDM, $m_H = 500\,\GeV{}$]{\includegraphics[width=0.5\textwidth]{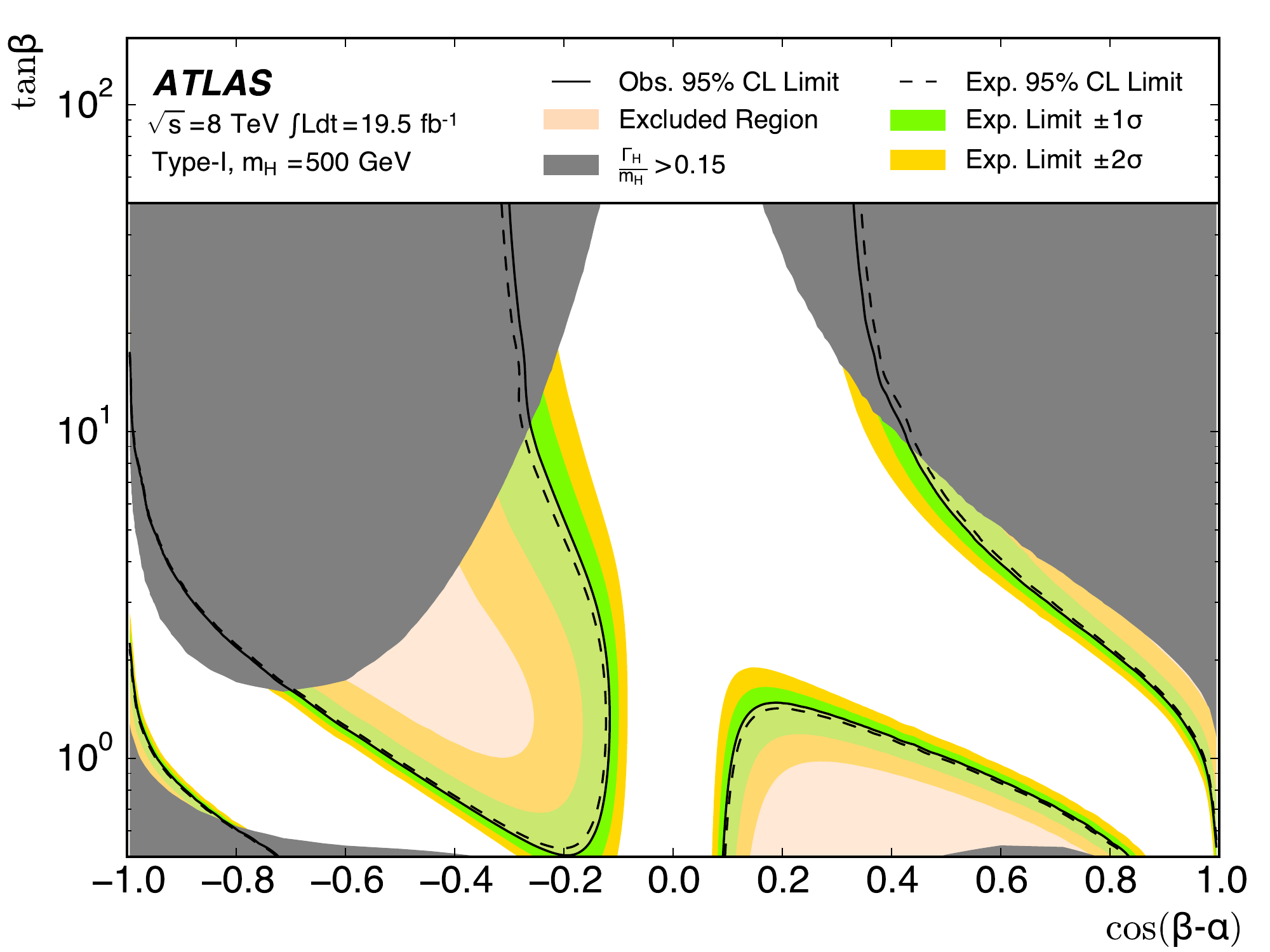}}
\subfloat[Type-II 2HDM, $m_H = 500\,\GeV{}$]{\includegraphics[width=0.5\textwidth]{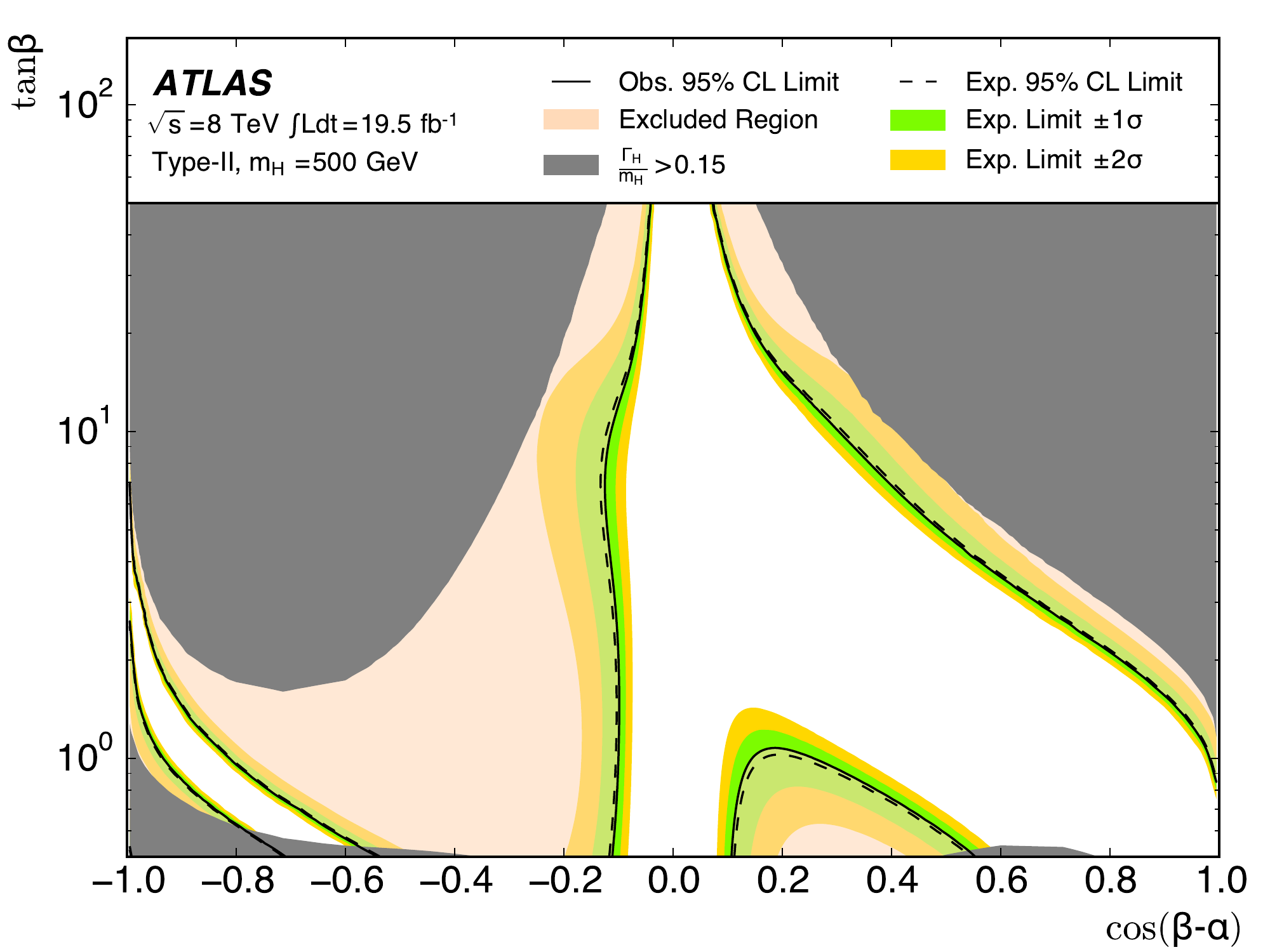}}
\caption{Excluded regions of the $\left(\cos\left(\beta - \alpha\right), \tan\beta\right)$ parameter space for (a) the Type-I 2HDM signal model and (b) the Type-II 2HDM signal model. The grey areas demarcate the phase-space regions where $\Gamma_H/m_H > 0.15$, for which the cross-section limits have not been demonstrated to be reliable.}
\label{fig:2HDM2Dt12}
\end{center}
\end{figure}

\begin{figure}[!ht]
\begin{center}
\subfloat[Lepton-specific 2HDM, $m_H = 500\,\GeV{}$]{\includegraphics[width=0.5\textwidth]{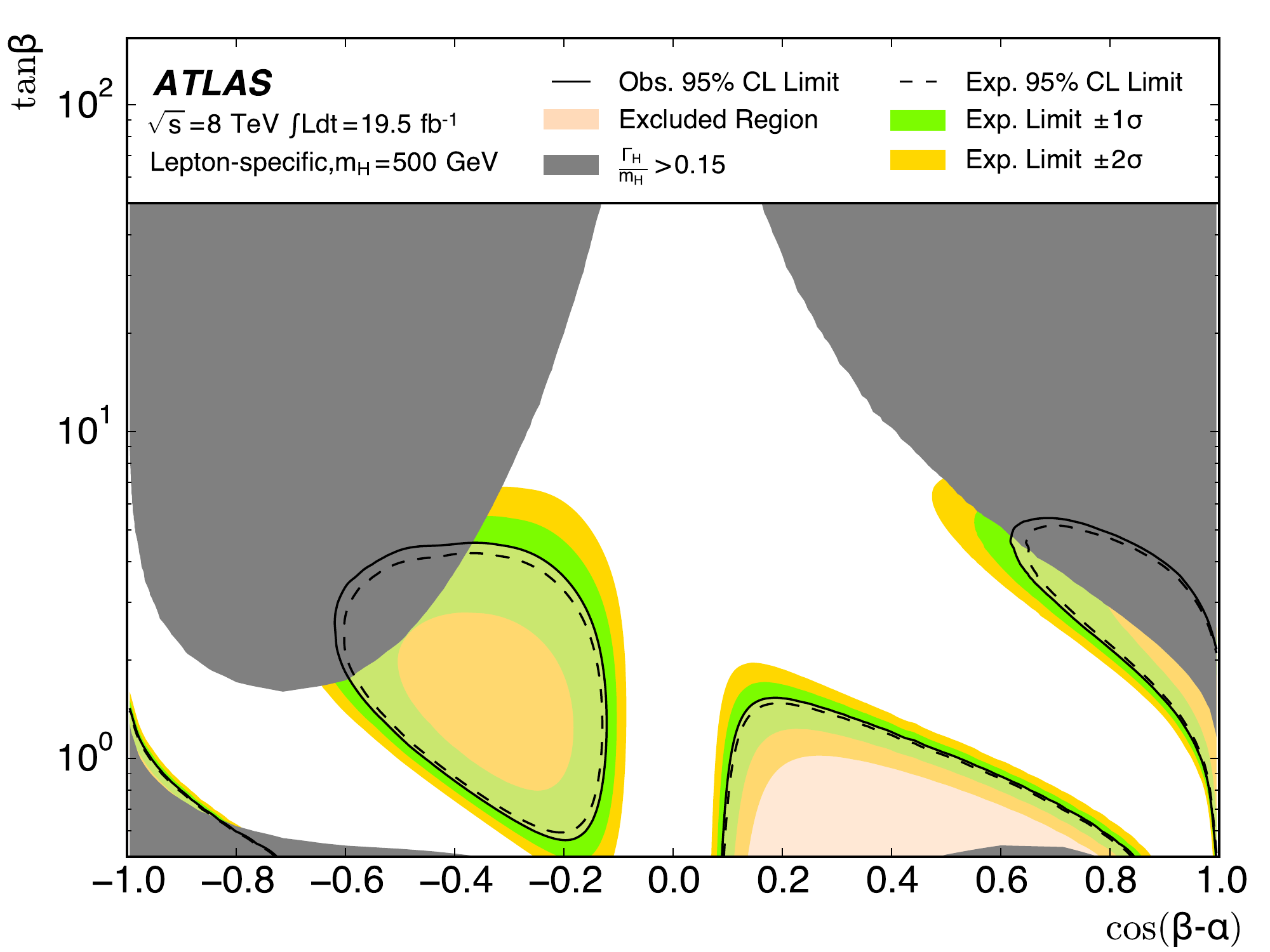}}
\subfloat[Flipped 2HDM, $m_H = 500\,\GeV{}$]{\includegraphics[width=0.5\textwidth]{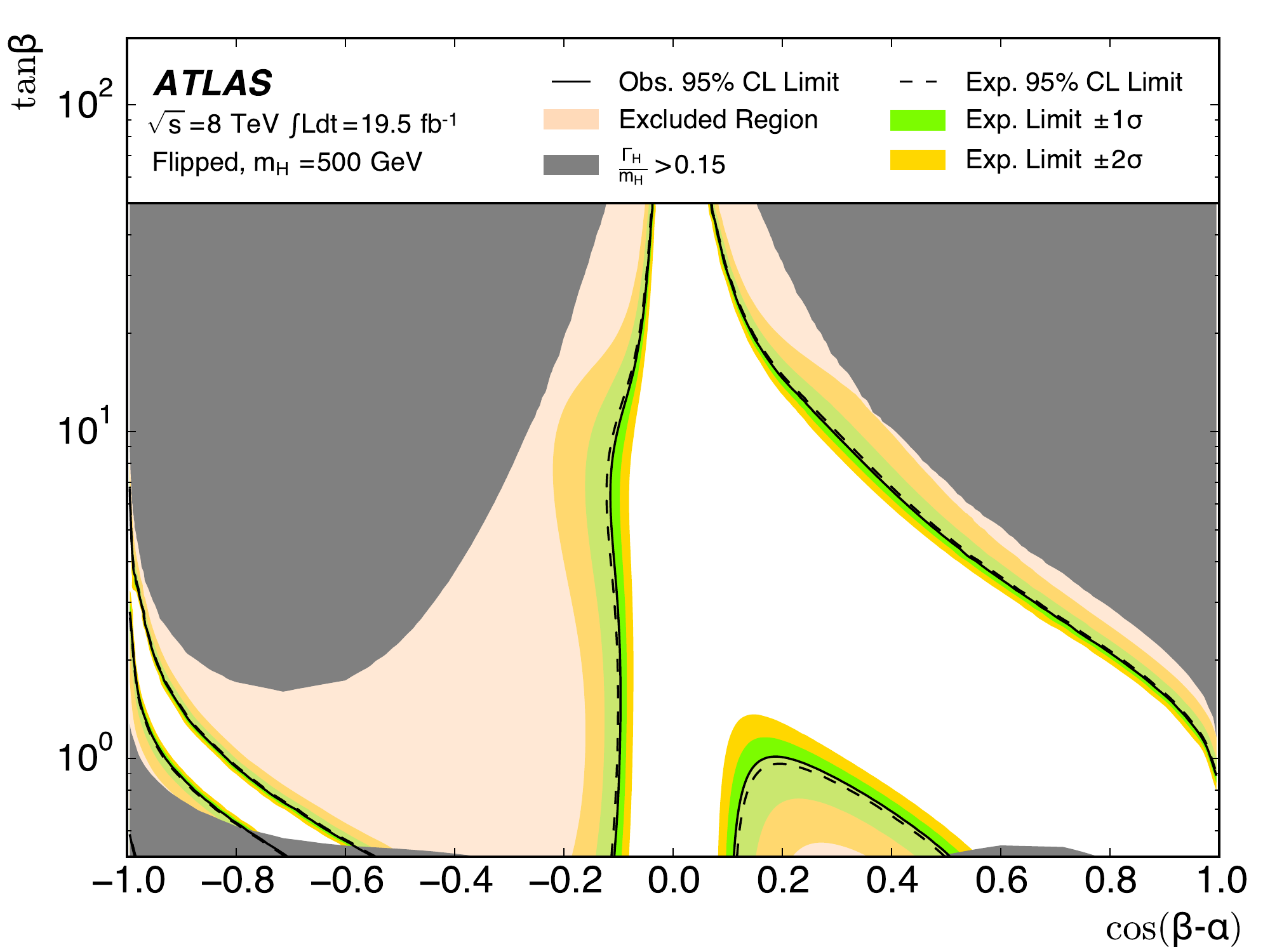}}
\caption{Excluded regions of the $\left(\cos\left(\beta - \alpha\right), \tan\beta\right)$ parameter space for (a) the Lepton-specific 2HDM signal model and (b) the Flipped 2HDM signal model. The grey areas demarcate the phase-space regions where $\Gamma_H/m_H > 0.15$, for which the cross-section limits have not been demonstrated to be reliable.}
\label{fig:2HDM2Dt34}
\end{center}
\end{figure}

\begin{figure}[!ht]
\begin{center}
\subfloat[Type-I 2HDM, $\cosba = -0.2$]{\includegraphics[width=0.5\textwidth]{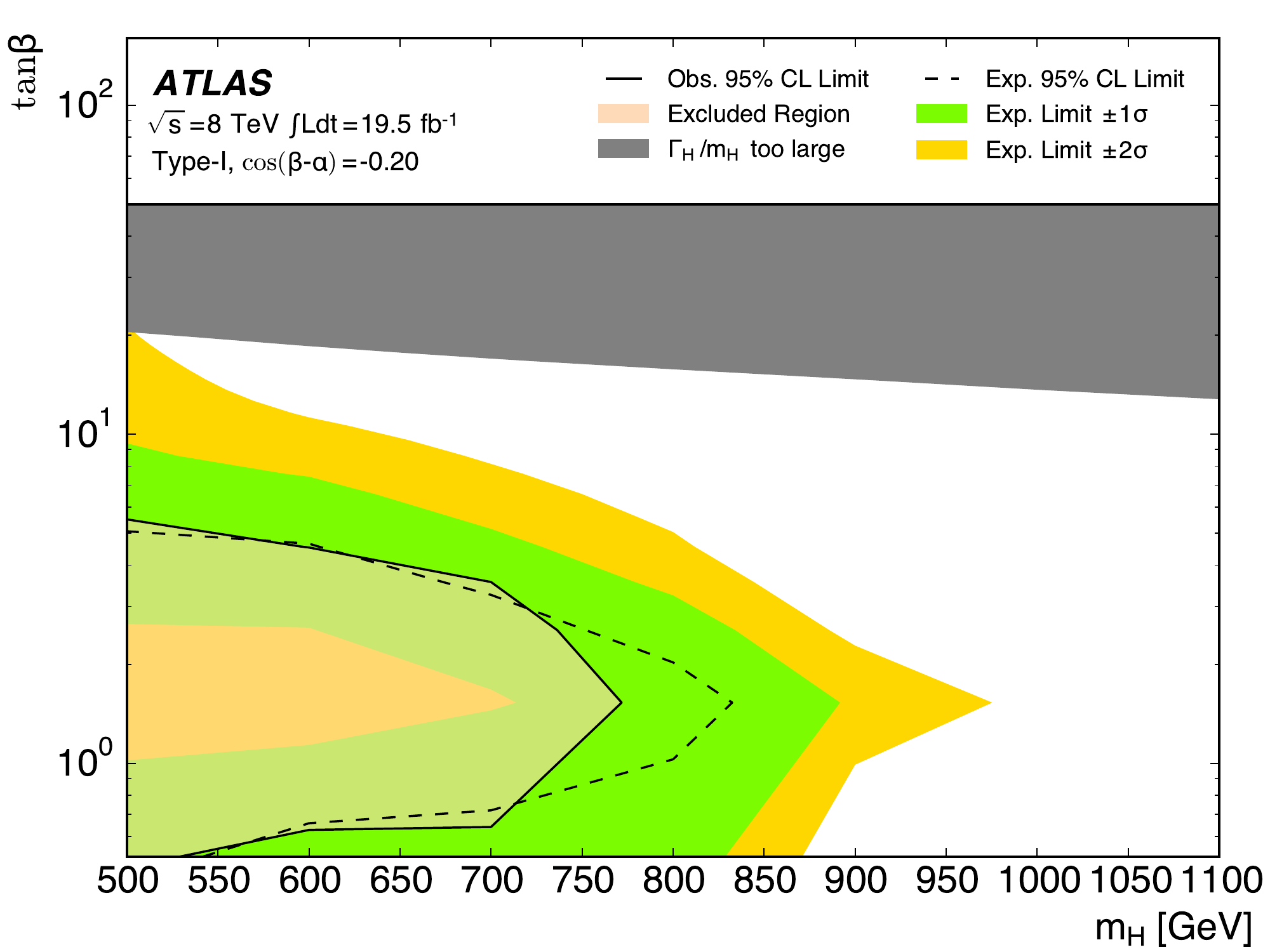}}
\subfloat[Type-II 2HDM, $\cosba = -0.2$]{\includegraphics[width=0.5\textwidth]{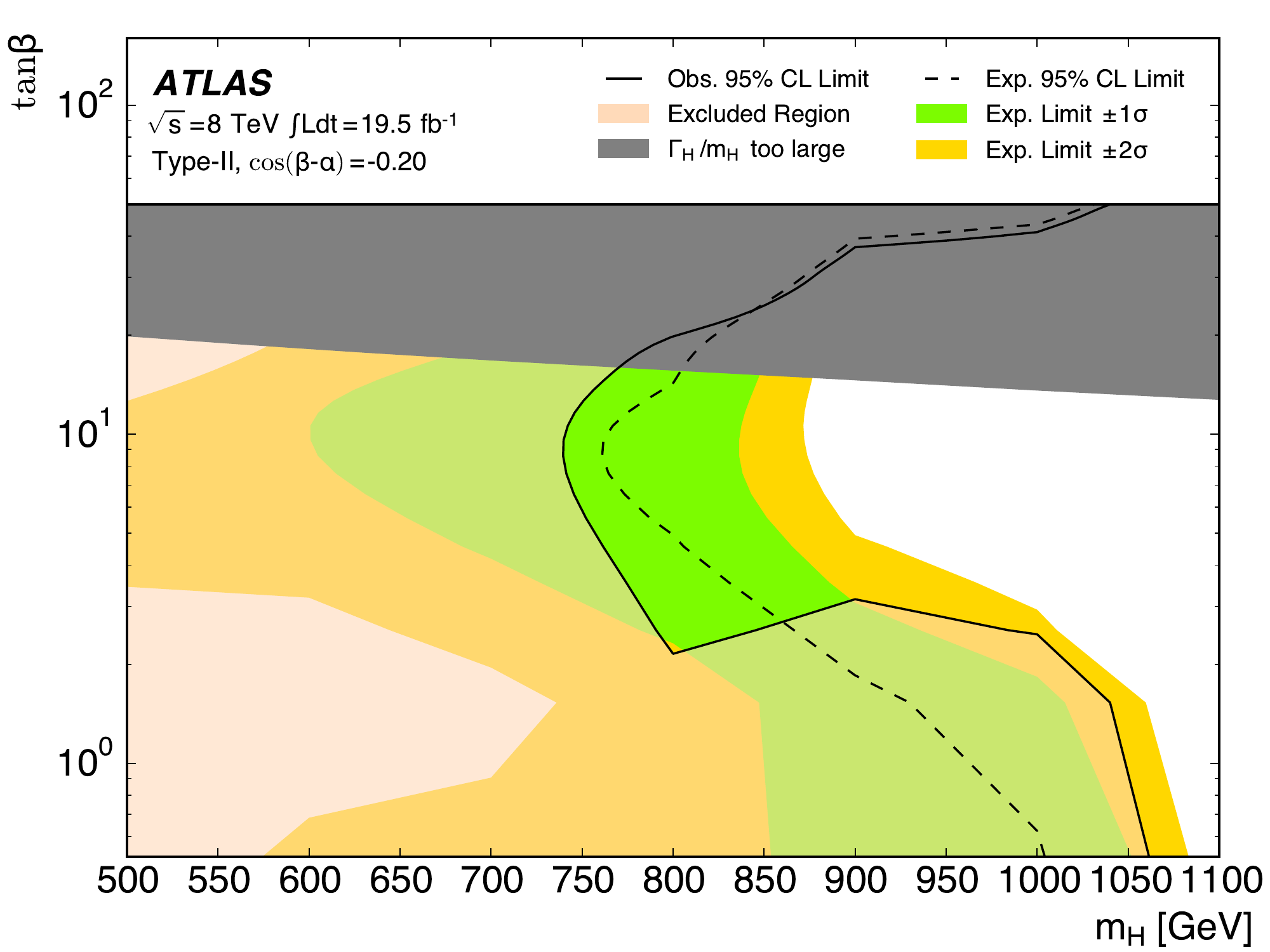}}
\caption{Excluded regions of the $\left(m_H, \tan\beta\right)$ parameter space for (a) the Type-I 2HDM signal model and (b) the Type-II 2HDM signal model. The grey areas demarcate the phase-space regions where $\Gamma_H/m_H$ is large ($\Gamma_H/m_H > 0.15$ for $m_H = 500\,\GeV{}$, increasing to $\Gamma_H/m_H > 0.23$ for $m_H = 1100\,\GeV{}$) and the limits have not been demonstrated to be reliable.}
\label{fig:2HDMmHtanbt12}
\end{center}
\end{figure}

\begin{figure}[!ht]
\begin{center}
\subfloat[Lepton-specific 2HDM, $\cosba = -0.2$]{\includegraphics[width=0.5\textwidth]{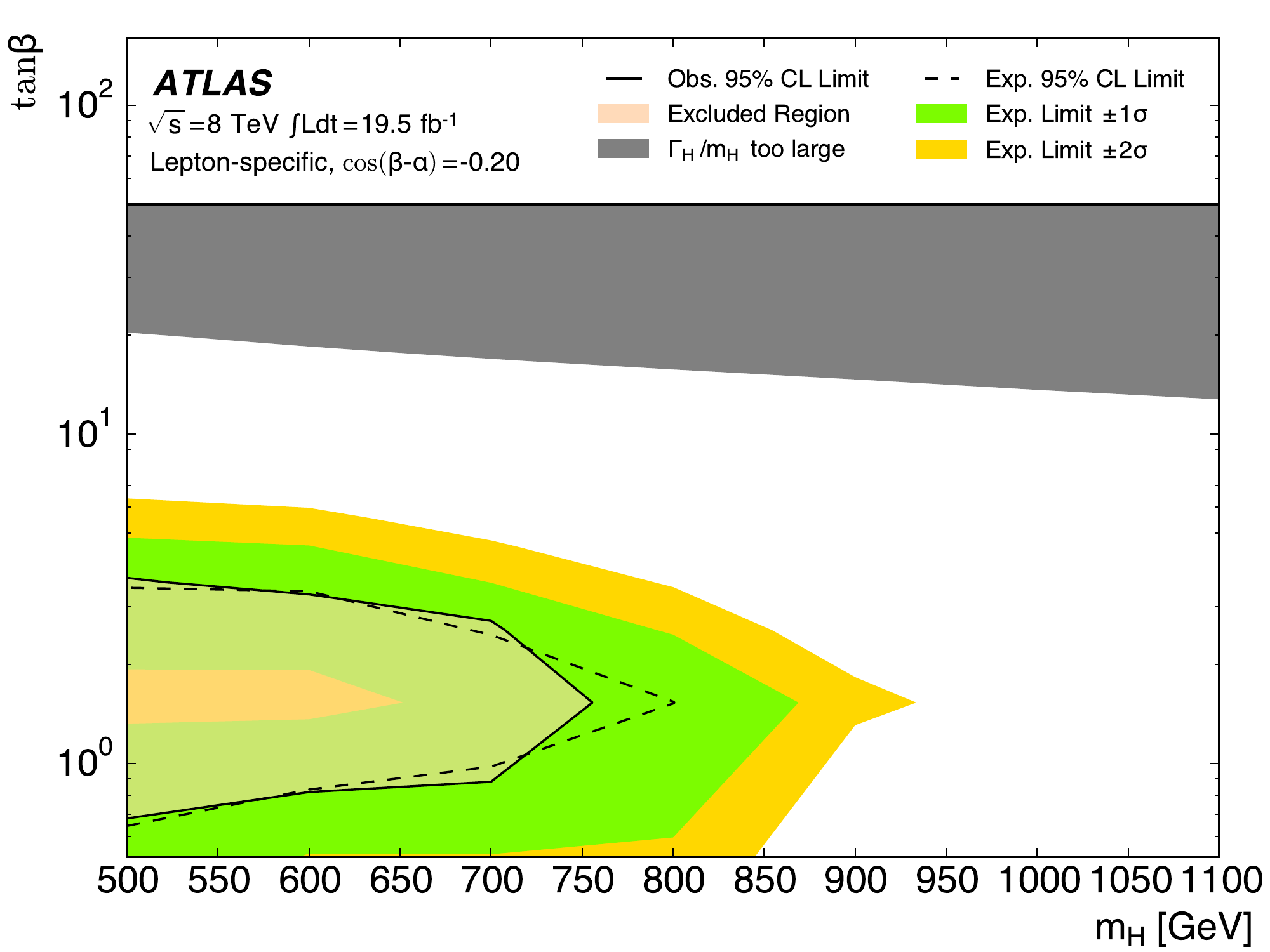}}
\subfloat[Flipped 2HDM, $\cosba = -0.2$]{\includegraphics[width=0.5\textwidth]{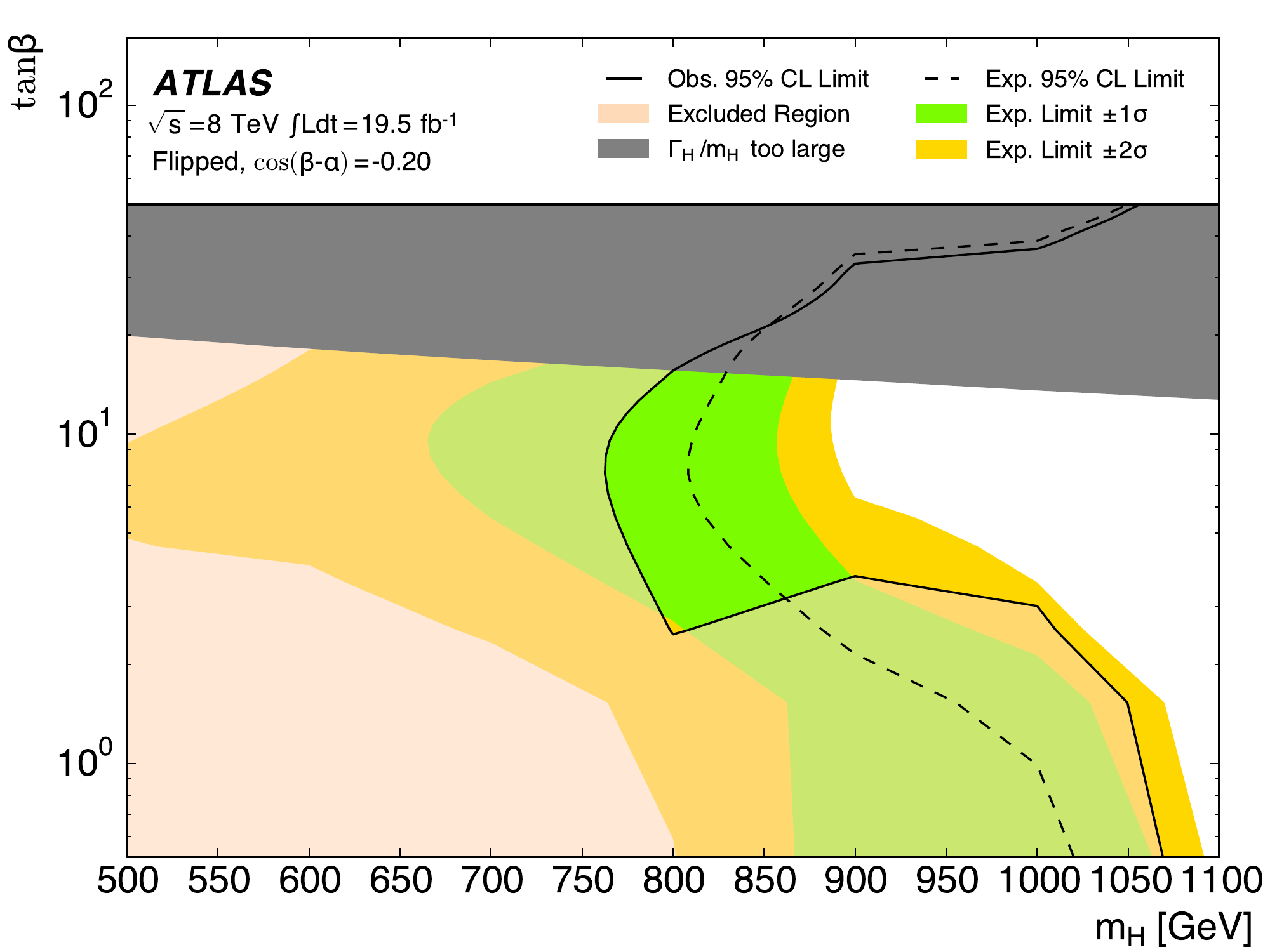}}
\caption{Excluded regions of the $\left(m_H, \tan\beta\right)$ parameter space for (a) the Lepton-specific 2HDM signal model and (b) the Flipped 2HDM signal model. The grey areas demarcate the phase-space regions where $\Gamma_H/m_H$ is large ($\Gamma_H/m_H > 0.15$ for $m_H = 500\,\GeV{}$, increasing to $\Gamma_H/m_H > 0.23$ for $m_H = 1100\,\GeV{}$) and the limits have not been demonstrated to be reliable.}
\label{fig:2HDMmHtanbt34}
\end{center}
\end{figure}

The effects of $\Gamma_H$ are accounted for by creating $m_H$ distributions with a range of widths, $0 < \Gamma_H/m_H \leq 0.5$, for each $m_H$ considered. These distributions are based on parameterizations which include resolution and acceptance effects combined with a Breit--Wigner line-shape. A grid of limits are calculated with each of these mass distributions. Then, for each point in $m_H$, \cosba, and \tanb space, the cross-section limit is determined by interpolating between the appropriate limits, based on the $\Gamma_H$ given by the model for that point. For the widest signals considered, the exclusion limits worsen by up to a factor of three. The exclusion regions determined through this process are shown as a function of \cosba and \tanb for $m_H=500\,\GeV{}$ in Figs.~\ref{fig:2HDM2Dt12} and \ref{fig:2HDM2Dt34}, and as a function of $m_H$ and \tanb for $\cosba = -0.2$ in Figs.~\ref{fig:2HDMmHtanbt12} and \ref{fig:2HDMmHtanbt34}. The validity of the process has been tested using the widest available signals, gravitons in the bulk RS model with \kMPl = 2. Phase-space regions with $\Gamma_H$ greater than these graviton widths are considered unvalidated and are shown in the figures as grey areas.



\label{sec:results}

\FloatBarrier

\section{Conclusions}
Two searches for Higgs boson pair production with the ATLAS detector at the LHC using the \fourb final state have been presented: one reconstructs Higgs boson candidates from pairs of nearby \akt $b$-tagged jets with $R = 0.4$;  the other reconstructs Higgs boson candidates using trimmed \akt jets with $R = 1.0$ matched to two $b$-tagged \akt track-jets with $R = 0.3$. Thanks to the high expected \hbb branching ratio and the large background rejection factors offered by the boosted dijet topology, the sensitivity for Higgs boson pair production is high, with a mass reach spanning the range between 500 and 2000~\GeV{}. There is no evidence for any signal in 19.5~\ifb of $pp$ collision data with $\sqrt{s} = 8$ \TeV. The largest deviation from the background-only hypothesis has a global significance of only 0.78~$\sigma$. The observed 95\% CL upper limit on \sigXfourb is 3.2 (2.3)~fb for narrow resonances with a mass of 1.0 (1.5) \TeV{}. 

Constraints are placed on several benchmark models. 
For the bulk RS model with $\kMPl=1$, KK gravitons in the mass range $500\leq$ \mGrav $\leq720$~\GeV{} are excluded at the 95\% CL. For non-resonant signals, using Standard Model $hh$ non-resonant production as the benchmark, the observed 95\% CL upper limit on $\sigma(pp\rightarrow hh\rightarrow b\bar{b}b\bar{b})$ is 202~fb, in good agreement with the expected exclusion.
This is to be compared to a SM prediction of $3.6 \pm 0.5$~fb.

\label{sec:conclusion}

%-------------------------------------------------------------------------------
\section*{Acknowledgements}
%-------------------------------------------------------------------------------

% Acknowledgements for papers with collision data
% Version 23-Mar-2015

% Standard acknowledgements start here
%----------------------------------------------
We thank CERN for the very successful operation of the LHC, as well as the
support staff from our institutions without whom ATLAS could not be
operated efficiently.

We acknowledge the support of ANPCyT, Argentina; YerPhI, Armenia; ARC,
Australia; BMWFW and FWF, Austria; ANAS, Azerbaijan; SSTC, Belarus; CNPq and FAPESP,
Brazil; NSERC, NRC and CFI, Canada; CERN; CONICYT, Chile; CAS, MOST and NSFC,
China; COLCIENCIAS, Colombia; MSMT CR, MPO CR and VSC CR, Czech Republic;
DNRF, DNSRC and Lundbeck Foundation, Denmark; EPLANET, ERC and NSRF, European Union;
IN2P3-CNRS, CEA-DSM/IRFU, France; GNSF, Georgia; BMBF, DFG, HGF, MPG and AvH
Foundation, Germany; GSRT and NSRF, Greece; RGC, Hong Kong SAR, China; ISF, MINERVA, GIF, I-CORE and Benoziyo Center, Israel; INFN, Italy; MEXT and JSPS, Japan; CNRST, Morocco; FOM and NWO, Netherlands; BRF and RCN, Norway; MNiSW and NCN, Poland; GRICES and FCT, Portugal; MNE/IFA, Romania; MES of Russia and NRC KI, Russian Federation; JINR; MSTD,
Serbia; MSSR, Slovakia; ARRS and MIZ\v{S}, Slovenia; DST/NRF, South Africa;
MINECO, Spain; SRC and Wallenberg Foundation, Sweden; SER, SNSF and Cantons of
Bern and Geneva, Switzerland; NSC, Taiwan; TAEK, Turkey; STFC, the Royal
Society and Leverhulme Trust, United Kingdom; DOE and NSF, United States of
America.

The crucial computing support from all WLCG partners is acknowledged
gratefully, in particular from CERN and the ATLAS Tier-1 facilities at
TRIUMF (Canada), NDGF (Denmark, Norway, Sweden), CC-IN2P3 (France),
KIT/GridKA (Germany), INFN-CNAF (Italy), NL-T1 (Netherlands), PIC (Spain),
ASGC (Taiwan), RAL (UK) and BNL (USA) and in the Tier-2 facilities
worldwide.
%----------------------------------------------


\clearpage
\pagebreak

%-------------------------------------------------------------------------------
% If you use biblatex and either biber or bibtex to process the bibliography 
% just say \printbibliography here
%\printbibliography
% If you want to use the traditional BibTeX you need to use the syntax below.
\bibliographystyle{atlasBibStyleWithTitle}
\bibliography{diboson4b_preprint}
%-------------------------------------------------------------------------------

\clearpage
\pagebreak

\newpage
% ATLAS Collaboration author list
% Data extracted on 16-Jul-2015 for paper reference EXOT-2014-11
\begin{flushleft}
{\Large The ATLAS Collaboration}

\bigskip

G.~Aad$^{\rm 85}$,
B.~Abbott$^{\rm 113}$,
J.~Abdallah$^{\rm 151}$,
O.~Abdinov$^{\rm 11}$,
R.~Aben$^{\rm 107}$,
M.~Abolins$^{\rm 90}$,
O.S.~AbouZeid$^{\rm 158}$,
H.~Abramowicz$^{\rm 153}$,
H.~Abreu$^{\rm 152}$,
R.~Abreu$^{\rm 30}$
% ...
\bigskip
\\
% ...
$^{11}$ Institute of Physics, Azerbaijan Academy of Sciences, Baku, Azerbaijan\\
% ...
$^{30}$ CERN, Geneva, Switzerland\\
% ...
$^{85}$ CPPM, Aix-Marseille Universit{\'e} and CNRS/IN2P3, Marseille, France\\
% ...
$^{90}$ Department of Physics and Astronomy, Michigan State University, East Lansing MI, United States of America\\
% ...
$^{107}$ Nikhef National Institute for Subatomic Physics and University of Amsterdam, Amsterdam, Netherlands\\
% ...
$^{113}$ Homer L. Dodge Department of Physics and Astronomy, University of Oklahoma, Norman OK, United States of America\\
% ...
$^{151}$ Institute of Physics, Academia Sinica, Taipei, Taiwan\\
$^{152}$ Department of Physics, Technion: Israel Institute of Technology, Haifa, Israel\\
$^{153}$ Raymond and Beverly Sackler School of Physics and Astronomy, Tel Aviv University, Tel Aviv, Israel\\
% ...
$^{158}$ Department of Physics, University of Toronto, Toronto ON, Canada\\
% ...
\end{flushleft}
% Created with ./xml2latex.py

\end{document}